\pdfoutput=1
\documentclass[%
 reprint,
 amsmath,amssymb,
 aps,
]{revtex4-2}
\usepackage{lipsum}
\usepackage{dcolumn}
\usepackage{color}
\usepackage{amsmath}
\usepackage{xcolor}
\usepackage{amsfonts}
\usepackage{comment}
\usepackage{braket}
\usepackage{esint}
\usepackage{bm}
\usepackage{cancel}
\usepackage{hyperref}
\usepackage{graphicx}
\usepackage{caption}
\usepackage{subcaption}
\usepackage[resetlabels,labeled]{multibib}
\usepackage[normalem]{ulem}
\newcommand{\bs}[1]{\boldsymbol{#1}}

\hypersetup{hypertex=true,
            colorlinks=true,
            linkcolor=blue,
            anchorcolor=blue,
            citecolor=blue}

\usepackage{enumerate}
\begin{document}

\preprint{APS/123-QED}

\title{A Replica-BCS theory for dirty superconductors}

\author{Yat Fan Lau}
\author{Tai Kai Ng}
\email{phtai@ust.hk}
\affiliation{%
Department of Physics, Hong Kong University of Science and Technology, Clear Water Bay Road, Kowloon, Hong Kong }%


\begin{abstract}
   
In this paper, we revisit the problem of dirty superconductors using a replica-symmetric BCS (RS-BCS) theory. We apply the RS-BCS theory to dirty superconductor grains of size $L^d$, where $L$ is the localization length and $d$ is the dimension of the system, assuming that the macroscopic system is composed of weakly coupled grains. Besides disordered potential, We also consider the case where regions with opposite signs of interaction exist in the grains, with net attractive interaction between electrons. Within the RS-BCS mean field theory, the system phase diagram, single-particle tunneling density of states and the superfluid density are computed within the RS-BCS theory for different strengths of disorder, We show that our result agrees qualitatively with previous numerical studies in the case with pure attractive interaction, and that a Cooper-pair-glass state may exist when regions with opposite sign of interaction exists and the net interaction between electrons is close to zero.  The plausible relevance of our result to the anomalous metal state is discussed.
\end{abstract}

\maketitle

\section{\label{sec:level1}Introduction\protect} 
Experimental observations in the past few decades reveal that an anomalous metal (or failed superconductor) state often exists between the superconductor-insulator (SIT) or superconductor-metal (SM) transition in a large number of superconductor thin film systems induced by change of some non-thermal parameters\cite{kapitulnik2019colloquium,Sacepa2020quantum,wang2023quantum}, and the challenge for physicists is to uncover the (presumably common) mechanism that leads to this surprising phenomenon. 
Motivated by these discoveries, we revisit the problem of dirty superconductors by developing a replica-symmetric BCS (RS-BCS) theory for the system, assuming that in addition to potential disorder, regions with opposite signs of interaction exist in the system with a net attractive interaction between electrons. We shall consider two dimensional (2d) systems or three dimensional (3d) systems with strong enough disorder such that all single-electron eigenstates are localized. In this case, the system can be viewed as consisting of weakly coupled dirty superconductor grains of size $\sim L^d$ where $L$ is the length of the electron localization and $d=$ the dimension of the system \cite{Sacepa2020quantum,bouadim2011single, read1995landau, 2023intrinsic}.  The RS-BCS theory is applied to describe a single dirty superconductor grain of size $\sim L^d$ with the assumption that the macroscopic system can be considered as dirty superconductor grains coupled weakly to each other.

Our approach generalizes the usual BCS mean-field theory by taking into account the plausible contributions from multiple solutions of the BCS mean-field theory in the presence of disorder and the presence of both attractive and repulsive regions in the system. It provides an efficient way of averaging over disorder compared with numerical calculations where a large number of systems with different disorder potential configurations have to be sampled\cite{bouadim2011single, ghosal2001inhomogeneous}. It also differs from other replica approaches which aim at deriving the field theoretical description(s) for the quantum superconductor-insulator(metal) transition or investigating other plausible fixed points starting from effective quantum rotor (or quantum $x-y$) models\cite{ioffe2010disorder,sknepnek2004order,feigel2000keldysh,phillips2003elusive,ye1993solvable,read1995landau}. The effect of a magnetic field that violates time-reversal symmetry and provides an explicit pair-breaking mechanism destroying superconductivity is not considered in this paper. 

 As we shall see in the following, our results agree qualitatively with previous analytic and numerical studies\cite{ma1985localized, kapitulnik1985anderson, bouadim2011single, ghosal2001inhomogeneous, granato2020disorder} when interaction between electrons is attractive and fixed. However, a Cooper pair glass state in which the phases and amplitudes of Cooper pairs are randomized is obtained when large, separate regions with different signs of interactions exist in the system with net interaction between electrons close to zero. The properties of the dirty superconductors are studied as a function of disorder strength and distribution of attractive and repulsive interactions in this paper. 
The plausible relevance of our findings to the anomalous metal state is discussed at the end of the paper. 

We start from a system of interacting electron gas with a general lattice Hamiltonian $H=H_0+H_\text{int}$, where \\\\
\begin{subequations}
\begin{align}
H_0 &= \sum_{i, j ; \sigma} t_{i j} c_{i \sigma}^{\dagger} c_{j \sigma}+\sum_{i \sigma} (W_i-\mu) n_{i \sigma}, \label{eq:1a} \\
H_{\mathrm{int}} &=  \sum_i U_i n_{i \uparrow} n_{i \downarrow}, \label{eq:1b}
\end{align}
where $c_i(c^\dagger_i)$ is the electron annihilation (creation) operator on site $i$ and $n_i=c^\dagger_ic_i$. $W_i$ is a random potential and $U_i$ is the onsite electron-electron interaction which can be positive or negative. We shall assume that regions with positive or negative electron interaction exist in the system of characteristic length $d_I$. The interaction $U_i$ is assumed to be the same within a region but fluctuates from region-to-region. The probability of a site $i$ being in the attractive[repulsive] region is $p[1-p]$. The mean interaction strength is $\braket{U_i}=-pU_0+(1-p)U_1$ where $U_0$ and $U_1$ are the interaction strengths in the attractive and repulsive regions, respectively.  The variance of interaction is
\begin{equation}
\label{Uvar}
\braket{U_iU_j}-\braket{U_i}^2=\sigma_U^2\theta(d_I-|\vec{x}_i-\vec{x}_j|). 
\end{equation}
\end{subequations}
where $\sigma_U=\sqrt{pU_0^2+(1-p)U_1^2-\braket{U_i}^2}=(U_0+U_1)\sqrt{p(1-p)}$.  $|\vec{x}_i-\vec{x}_j|$ is the distance between two sites $i,j$. We assume that strong potential barriers between regions with different signs of interaction {\em do not exist} in our model and the electron eigenstates are NOT localized within regions with fixed interaction as is usually assumed for granular materials\cite{beloborodov2007granular}.

In the eigenstate basis given by $[c_{k\sigma},H_0-\mu N]=\xi_k c_{k\sigma}$
where $N$ is the total particle number operator and $c_{k\sigma}=\sum_i\phi_{ki}c_{i\sigma}$ where $\phi_{ki}$ are the eigenstates of $H_0$, (we set $\hbar=1$), the Hamiltonian can be written as(see also \cite{ghosal2001inhomogeneous})
\begin{equation}
H=\sum_{k \sigma} \xi_k c_{k \sigma}^{\dagger} c_{k \sigma}+\sum_{k l, p q} U_{k l, p q} c_{k \uparrow}^{\dagger} c_{l \downarrow}^{\dagger} c_{p \downarrow} c_{q \uparrow},
\end{equation}
where $\xi_k$ is the single-particle energy of state $k$ measured from the chemical potential $\mu$ and
\[ U_{k l, p q}= \sum_i U_i \phi^*_{ki} \phi^*_{li} \phi_{pi}\phi_{ki}. \] 
 We note that $k$ represents eigenstates of $H_0$ which are not plane-wave states. We shall assume that $\xi_k$ forms a continuous spectrum across the Fermi surface and denote the time-reversal state of state $k$ by $-k$, with $\xi_k=\xi_{-k}$ in the presence of time-reversal symmetry.  A mean-field BCS decoupling of $U_{k l, p q}$ is performed, with
\begin{widetext}
\begin{equation}
\label{mfd}
\sum_{klpq}U_{k l, p q} c_{k \uparrow}^{\dagger} c_{l \downarrow}^{\dagger} c_{p \downarrow} c_{q \uparrow} \sim \sum_{k\neq p}U_{k p} \left( c_{k \uparrow}^{\dagger} c_{-k \downarrow}^{\dagger} \langle c_{p \downarrow} c_{-p \uparrow}\rangle  + \langle c_{k \uparrow}^{\dagger} c_{-k \downarrow}^{\dagger}\rangle c_{p \downarrow} c_{-p \uparrow}-\langle c_{k \uparrow}^{\dagger} c_{-k \downarrow}^{\dagger}\rangle \langle c_{p \downarrow} c_{-p \uparrow}\rangle \right)  + \sum_kU_{k k} c_{k \uparrow}^{\dagger} c_{-k \downarrow}^{\dagger} c_{-k \downarrow} c_{k \uparrow} 
\end{equation}
\end{widetext}
where $U_{k p}=\sum_i U_i\left|\phi_{ki}\right|^2\left|\phi_{pi}\right|^2$, i.e., we follow Anderson and assume that a BCS wavefunction for dirty superconductors can be constructed by pairing the time-reversal eigenstates of the single-particle Hamiltonian\cite{anderson1959theory, ma1985localized}. We shall see that the effect of $U_{kk}$ is qualitatively different from $U_{kp} (p\neq k)$ for localized wavefunctions and should be treated separately as we shall discuss in the following and in Appendix \ref{ME}. 

With this decoupling, the mean-field ground state energy of the system is characterized by an effective energy
\begin{subequations}
\begin{equation}
\label{E0}
\begin{aligned}  &\quad E(\Delta,\Delta^*;\lambda,\lambda^*) \\
&=-\sum_k E_k+\sum_{k\neq p}U_{kp}\Delta_k^*\Delta_p+\sum_k(\lambda_k\Delta_k^*+\lambda_k^*\Delta_k) 
\end{aligned}
\end{equation}
where $E_k=\sqrt{\xi_k^2+|\lambda_k|^2}$. The ground state energy of the system is obtained by minimizing $E(\Delta,\Delta^*;\lambda,\lambda^*)$ with respect to $\lambda_k$'s and $\Delta_k$'s, where we obtain the usual BCS mean-field equations
\begin{equation}
\label{bcsmf}
    \lambda_k=-\sum_{p\neq k} U_{k p}\Delta_p, \quad \quad \Delta_p=\frac{\lambda_p}{2E_p}.
\end{equation}
 The calculation can be further simplified by introducing unit vector fields $\vec{s}_k$ defined by
 \begin{equation}
 \label{ss}
 s^{z}_k  = \frac{\xi_k}{E_k}  \quad \quad s^{x}_k+is^{y}_k = \frac{\lambda_k}{E_k}.
\end{equation}
 Using Eqs.\ (\ref{E0}) and\ (\ref{bcsmf}), we obtain $E(\Delta,\Delta^*;\lambda,\lambda^*)\rightarrow E(\{\vec{s}\})$, where
 \begin{equation}
\label{Eeff}
   E(\{\vec{s}\})=\sum_k -\xi_ks^{z}_k+\frac{1}{4}\sum_{k\neq p}U_{kp}\left(s^{x}_ks^{x}_p+s^{y}_ks^{y}_p\right).
\end{equation}
\end{subequations}
The ground state energy is obtained by minimizing $E(\{\vec{s}\})$ with respect to $\vec{s}_k$'s. Notice that the effect of disorder is completely encoded in $U_{kp}$ in our approach. We shall show in Appendix \ref{ME} that $U_{kk}$ does not contribute to $E(\{\vec{s}\})$ but contributes to the single-particle energy gap.

For disordered systems, we have to compute the average effective energy $\langle E(\{\vec{s}\})\rangle_d$ and also $\langle A\rangle_d$ for observable $A$'s where $\langle...\rangle_d$ denotes average over disorder. We shall employ the Replica trick\cite{sherrington1975solvable} to perform the disorder-average. Before introducing the replica trick, we first describe what we expect on physical grounds.

\begin{figure}
\centering
\includegraphics[width=\columnwidth]{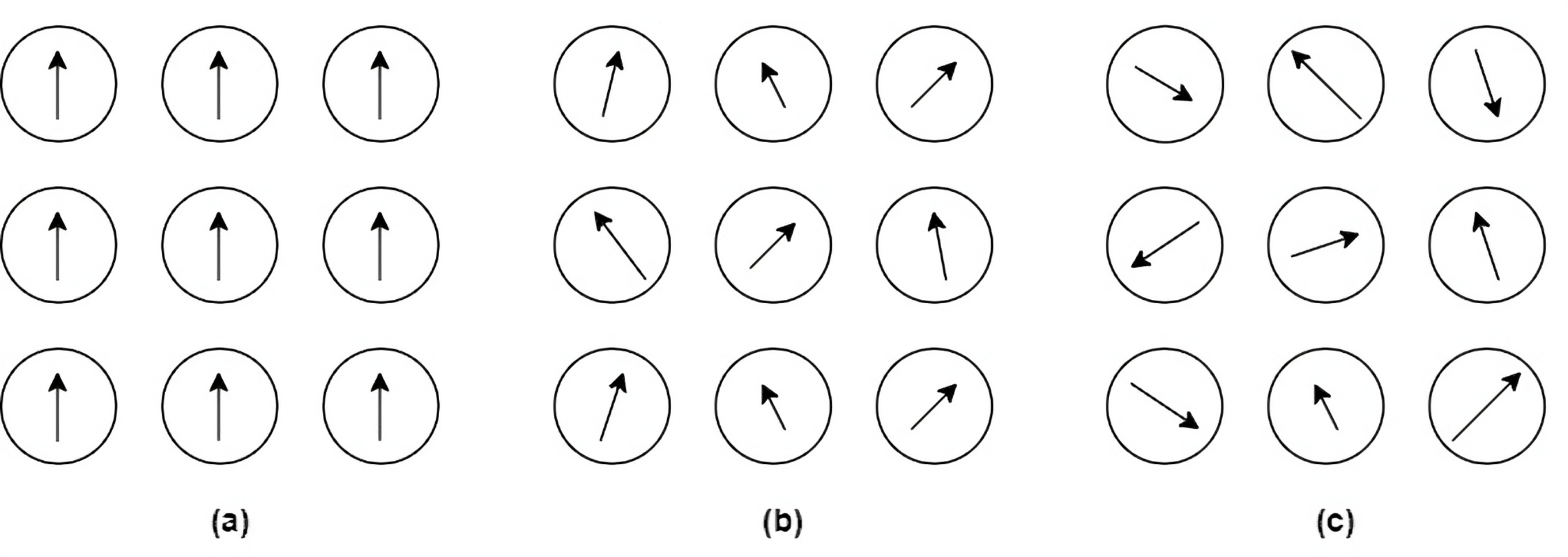}
\caption{A schematic diagram of the order parameters $\Delta_k$ for different disorder strengths}
\label{spinor}
\end{figure}

From the mean-field equation Eq.(\ref{bcsmf}), we see that when the interaction matrix element $U_{kp}$ is not a constant, the mean-field gap function $\lambda_k$'s are in general different for different $k$'s. The situation is represented in Fig.\ref{spinor}, where the arrows in each circle denote $\lambda_k$  for a particular value of $k$, the magnitude of the arrow representing $|\lambda_k|$, and the direction representing the phase angle $\lambda_k=|\lambda_k|e^{i\theta_k}$.

In the limit $U_{kp}=-U_0$, it is easy to show from Eq.\ (\ref{bcsmf}) that $\lambda_k=\lambda_0$ are all equal and the situation is represented in Fig.\ref{spinor}a. For $U_{kp}=-U_0+\delta U_{pk}$ where $\delta U_{pk}$ represents a small, random fluctuating part of $U_{kp}$, we expect small fluctuations also exist in $\lambda_k$ and the result is represented in Fig.\ref{spinor}b, where $\lambda_k$'s are different for different $k$ but still has a nonzero average value. We shall call this a dirty superconductor state. For strong enough fluctuations $\delta U_{pk}$, the average value of $\lambda_k$ may vanish (Fig.\ref{spinor}c) and we shall call this the Cooper-pair-glass state (CPG). The Cooper pair-glass state is qualitatively similar to the phase-glass state obtained in the random quantum rotor model\cite{ioffe2010disorder,sknepnek2004order,feigel2000keldysh,phillips2003elusive,ye1993solvable,read1995landau,granato2020disorder}  or the Bose-glass state\cite{fisher1989boson, weichman2008particle} except that both the amplitude and phases of Cooper pairs fluctuate in the present case. We note that the CPG state is expected to occur {\em only} when both negative and positive interaction regions exist and the system is frustrated. Multiple solutions to the BCS mean-field equation are expected to contribute to $E(\{\vec{s}\})$ and become important in this case\cite{mezard1987spin,nishimori2001statistical,binder1986spin}. 

We caution that $\lambda_k$'s represent static mean-field solutions of BCS theory which is insufficient when quantum fluctuation is strong and the system is quantum disordered\cite{mpafisher1986,fisher_1990,fazio1991charge, Sacepa2020quantum}. The effects of quantum fluctuations will be discussed at the end of the paper.  We note also that the BCS mean-field theory can be extended to finite temperature straightforwardly by replacing BCS ground state energy by the corresponding free energy. We shall restrict ourselves to temperature $T=0$ in this paper.

\section{\label{sec:level1}the Replica approach\protect} 
   In this section we explain how we apply the replica approach to compute the disorder-averaged effective energy $\langle E(\{\vec{s}\})\rangle_d$ and observable $\langle A\rangle_d$'s. We shall explain the approximations we made, the limitations of the approach and the main results we obtain. Mathematical details of our calculations are presented in Appendices \ref{ME} and \ref{MFE}.

   The main assumption we made in our approach is that for continuous spectrum $\xi_k$ across the Fermi surface, the major effect of disorder on the effective energy (\ref{Eeff}) appears in the interaction matrix element $U_{pk}$. To implement the replica approach, we shall neglect the correlation between different matrix elements $U_{pk}, U_{p'k'}$ and assume that $U_{pk}$ follows a simple $(k,p)$-independent Gaussian distribution
 \begin{equation}
    P(U_{kp}) \sim \sqrt{\frac{1}{2\pi\sigma^2}} \exp{\left({-\frac{(U_{kp}-U_m)^2}{2\sigma^2}}\right)}
\label{Gaussian}
\end{equation}
where $U_m\sim \frac{\braket{U_i}}{\mathcal{V}}$ represents an average attractive interaction between states $k$ and $p$, $\sigma^2$ is the corresponding variance. $\mathcal{V}$ is the volume of the system we consider here. We shall write $\sigma^2=\frac{h^2}{\mathcal{V}}$ for technical convenience (see Appendix B). We note that Eq.\ (\ref{Gaussian}) is a very crude approximation as {\em all properties of the dirty superconductor are characterized by only two parameters $U_m$ and $\sigma^2$}. In particular, the precise spatial information is lost when we assume that $P(U_{kp})$ is $k,p$-independent, as the approximation corresponds to infinite-range interaction in $k$-space. The approximation is qualitatively valid when we apply it to a grain of size $\mathcal{V}\sim L^d$, where all states in the grain remain strongly coupled to each other but not to states with distance $\gg L$ from each other (see also Appendix \ref{ME}). The purpose of our study is to find out to what extent the properties of dirty superconductors can be recovered by this crude but simple approximation for superconductors with pure attractive interaction and to see what it suggests for dirty superconductors with regions of both attractive and repulsive interactions. 

 We have estimated $U_m$ and $\sigma^2$ for localized single-particle wavefunctions for the two situations discussed in the {\em Introduction}: (1)$U_i=-U_0(U_0>0)$ for all lattice sites $i$, and (2) regions of size $d_I$ with different interactions $-U_0$ and $U_1$ exist in the system. We assume that the electronic wavefunctions are insensitive to the boundaries between different regions when we estimate $U_m$ and $\sigma^2$. Writing 
\[
\left|\phi_k(\bs{x})\right|^2 \sim \frac{1}{L^d} e^{-\frac{\left|\bs{x}-\bs{X}_k\right|}{L}}
 \]
where $L$ is the localization length, $d$ the spatial dimension and $\bs{X}_k$ the center of localization\cite{2023intrinsic}) which is randomly located in the system, we find that in case (1) with $k\neq p$,
\begin{subequations}
\label{h&L1}
\begin{equation}
\langle U_i\rangle \sim -U_0,
\end{equation}
and
\begin{equation}
\label{uv21}
h^2=L^d\times\left(\braket{U_{kp}^2}_{d}-\braket{U_{kp}}_{d}^2\right) \sim \frac{U_0^2}{L^{d}}.
\end{equation}
\end{subequations}
 (notice we set $\mathcal{V}=L^d$) whereas in case (2)
\begin{subequations}
\label{h&L2}
\begin{equation}
   \braket{U_i}\sim (-pU_0+(1-p)U_1),
\end{equation}
and
\begin{equation}
\label{uv22}
h^2 \sim \frac{\sigma_U^2 D_I^d}{L^{2d}} + \frac{\braket{U_i}^2}{L^{d}}
\end{equation}
\end{subequations}
where $D_I=d_I$ for $d_I \leq L$ and $D_I=L$ for $d_I > L$. Notice that for $d_I>L$, the whole grain is essentially located in a region with fixed interaction $-U_0$ or $U_1$ and our approximation for $\braket{U_i}$ breaks down. The detailed derivations are presented in Appendix \ref{ME}. We note that $h^2$ vanishes when $L\rightarrow\infty$, indicating that our approximation is not applicable to describe weakly-disordered 3D dirty superconductors where electronic eigenstates remain extended and a more refined approach is needed\cite{ma1985localized, kapitulnik1985anderson}.

 For $d_I \ll L$, $h^2$ is mainly determined by the term $\braket{U_i}^2/L^{d}$ which reflects the effect of disorder on electronic wavefunctions (see Appendix \ref{ME}). The term $\frac{\sigma_U^2 D_I^d}{L^{2d}}$ is a result of the existence of regions of opposite signs of interaction.  We shall see that this term leads to plausible existence of Cooper-pair-glass state when $\braket{U_i}$ is small enough and $d_I/L\sim O(1)$.

To apply the replica trick we write $E(\{\vec{s}\})=F(\{\vec{s}\};T\rightarrow 0)=-\beta^{-1}\ln{Z(T \to 0)}$ where $T=$ temperature and $Z(T) = \text{Tr}e^{-\beta E(\{\vec{s}\})}$, $\beta=(k_BT)^{-1}$. The disorder-averaged ground state energy is then given by
\begin{equation}
\begin{aligned}
\label{DP}
   \braket{E_G}_d =\lim_{T\rightarrow0}\int D\left[U_{kp}\right] P\left(\left[U_{kp}\right]\right) \left(-\frac{1}{\beta}\ln{Z\left(T;\left[U_{kp}\right]\right)}\right)
\end{aligned}
\end{equation}
where $D\left[U_{k p}\right]=\prod_{k p} d U_{k p},  P\left(\left[U_{k p}\right]\right)=\prod_{k p} P\left(U_{k p}\right)$ and $Z\left(T;\left[U_{kp}\right]\right)$ is the partition function for a particular configuration of $\{U_{kp}\}$. To compute $\braket{\ln Z}_d$ we employ the mathematical identity (replica trick)
\begin{equation}
    \braket{\ln{Z}}_d = \lim_{n \to 0} \frac{\braket{Z^n}_d-1}{n}.
    \label{limit}
\end{equation}
where we compute $\braket{Z^n}_d$ for $n=$ integer and then analytically continue the result to the $n \to 0$ limit. 

$\braket{Z^n}_d$ can be computed easily with Eqs.\ (\ref{Gaussian}) and\ (\ref{DP}). We note that with Eq.\ (\ref{Gaussian}) for $P(U_{kp})$ which is independent of $(k,p)$,
the resulting expression for $\braket{Z^n}_d$  can be treated by a standard replica-mean-field theory\cite{sherrington1975solvable,mezard1987spin,nishimori2001statistical,binder1986spin}. We consider the replica-symmetric mean-field solution in this paper and have checked that the solution is stable with respect to the replica symmetry breaking terms at zero temperature. The stability analysis is detailed in Appendix \ref{Stability}. We shall summarize the outcome of the mean-field theory in this section.

The disorder-averaged ground state energy is given by
\begin{widetext}
\begin{equation}
\begin{aligned}
\braket{E_G}_d = -\sum_k \int \frac{d \bs{y_{k}}}{2 \pi} \exp{\left[-\frac{\left(\bs{y_{k}}\right)^2}{2}\right]} \frac{1}{\beta} \ln \left(Z_{0 k}(\bs{y_{k}})\right)+\frac{\Delta^2}{\bar{U}}+2 \varphi_{+} \varphi_{-}
\end{aligned}
\end{equation}
\end{widetext}
where $\bs{y_k}=(y_{kx},y_{ky})$, $\bar{U}=|\braket{U_i}|$,
$d\bs{y_k}=d y_{kx}d y_{ky}$, $Z_{0 k}(\bs{y_{k}})=\int D [\vec{s}_k] e^{-\beta f_{eff}(\vec{s}_k,\bs{y_{k}})}$ with
\begin{widetext}
\begin{equation}
\label{feff}
f_{eff}(\vec{s_k},\bs{y_{k}})=\xi_k s_k^{(z)}-\bs{s_{k}^{ \perp}} \cdot\left(\sqrt{\frac{h \varphi_{+}}{2}} \bs{y_k}+\Delta \bs{\hat{x}}\right)-\frac{1}{4}h \varphi_{-}(\bs{s_{k}^{ \perp}})^2,
\end{equation}
\end{widetext}
where $\bs{s^\perp_k}=(s_k^{(x)},s_k^{(y)})$. Three mean-field order parameters are obtained in the replica-symmetric mean-field theory (see Appendix \ref{MFE} for details) including the average superconductor order-parameter $\Delta$, the Cooper-pair-glass order-parameter $h\varphi_-$ and the Cooper-pairs amplitude parameter $h\varphi_+$ which are determined by the following self-consistent mean field equations at $T=0$,
\begin{equation}
\label{mfeq}
\begin{gathered}
h\varphi_{-}=\frac{h}{4 \varphi_+} \sqrt{\frac{h \varphi_+}{2}}\frac{1}{\mathcal{V}} \sum_k\left\langle \braket{\bf{s_{k}^{\perp}} \cdot \bs{y_{k}}}\right\rangle_d \\
h\varphi_{+}=\frac{h^2}{8}\frac{1}{\mathcal{V}} \sum_k \braket{ \braket{ \bs{s_{k}^{\perp }} \cdot \bs{s_{k}^{\perp }}} }_d  \\
\Delta=\frac{\bar{U}}{2}\frac{1}{\mathcal{V}}\sum_k \braket{\left\langle s_{k x}\right\rangle}_d
\end{gathered}
\end{equation}
where $\braket{\braket{A(\bs{\vec{s}_k})}}_d$ consists of two averages given by
\begin{subequations}
\label{observable}
\begin{equation}
\begin{aligned}
\braket{\braket{A(\vec{s}_{k})}}_d =\int \frac{d \bs{y_{k}}}{2 \pi} \exp{\left[-\frac{\left(\bs{y_{k}}\right)^2}{2}\right]}\braket{A(\vec{s}_{k},\bs{y_{k}})}
\end{aligned}
\end{equation}
and
\begin{equation}
\braket{A(\vec{s}_{k},\bs{y_{k}})} =\left(\frac{1}{Z_{0 k}(\bs{y_{k}})}\right)\int D[\vec{s}_k] A(\vec{s}_{k})e^{-\beta f_{eff}(\vec{s}_k,\bs{y_{k}})}.
\end{equation}
\end{subequations}
 $\braket{A(\vec{s}_{k},\bs{y_{k}})}$ is the thermal average over an effective $k$-dependent free energy in Eq. (\ref{feff}). At $\beta \to \infty$, this average can be computed in the saddle point approximation where we replace $\braket{A(\vec{s}_{k},\bs{y_{k}})}$ by its saddle point value $A(\vec{s}_{k}^{(m)},\bs{y_{k}})$ where $\vec{s}_k^{(m)}$ minimizes $f_{eff}(\vec{s}_k,\bs{y_{k}})$. The BCS mean-field theory is recovered in this step (see Eq.\ (\ref{Eeff}) and discussion thereafter). The second bracket $\braket{A}_d=\int \frac{d \bs{y_{k}}}{2 \pi} A(\bs{y_{k}}) e^{-\frac{\left(\bs{y_{k}}\right)^2}{2}}$ denotes an average over configurations $\bs{y_k}$ in $f_{eff}(\vec{s}_k,\bs{y_{k}})$.

Physically, we can interpret $\bs{B}_k=\sqrt{\frac{h \varphi_{+}}{2}} \bs{y_k}+\Delta\bs{\hat{x}}$ as an effective $\bs{y_k}$-dependent pairing field coupling to each Cooper pair $k$ and the replica-symmetric mean-field theory introduces essentially two approximations: 
(1) it replaces the energy $E(\{\vec{s}\})$ by effective $k$-dependent free energies $f_{eff}(\vec{s_k},\bs{y_{k}})$'s where coupling between different $k$-states is replaced by coupling of individual $k$-states to effective pairing fields $\bs{B}_k$'s and 
(2) the different disorder configurations lead to a random component $\bs{y_k}$ in $\bs{B}_k$ that follows a Gaussian distribution. $f_{eff}(\vec{s_k},\bs{y_{k}})$ and the coupling of Cooper pairs to $\bs{B}_k$ are characterized by the three mean-field parameters $\Delta$, $h\varphi_+$ and $h\varphi_-$ which are determined self-consistently.

The order parameters $\Delta$, $h\varphi_-$ and $h\varphi_+$ have the following physical meaning: $\Delta$ is the usual superconductivity order parameter for a dirty superconducting grain, $h\varphi_-$ is the glass order parameter, corresponding to the Edward-Anderson order parameter $q_{EA}$ in spin-glasses\cite{mezard1987spin,nishimori2001statistical,binder1986spin}. $h\varphi_+$ is the pseudo-gap order parameter measuring the average magnitude of the local superconducting gap. This parameter is absent in spin glass or $x-y$ models where the spin magnitude is fixed. 

\section{\label{sec:level1}Mean-field calculation results\protect} 

We have solved the mean-field equations Eq.(\ref{mfeq}) numerically and the results of our calculations are presented in this section. We choose $N(0)U_0=0.5, U_0=0.2 eV$, $d_I=15a$ ($a=$ lattice spacing), $\omega_D=20meV$,  where $N(0)$ is the density of states on Fermi surface and $\omega_D$ is the Debye frequency in presenting our results. We assume that these parameters are not affected by disorder in our calculations and apply our analysis to two dimensions where all states are localized. 

The results of our mean field calculation are presented as a function of $L^{-1}$ for two cases: (i) $U_i=-U_0$ for all sites $i$, and (ii) regions with different signs of interaction exist, with $p=0.55$ and $U_1=U_0$. Notice that the mean interaction $\braket{U_i}$ and the fluctuation $h^2$ are determined by four parameters: $p$, $U_0$, $U_1$ and $d_I$ and different combinations of these parameters could give the same results. For example, with the same $U_0$ and $d_I$, both $p=0.8,U_1=U_0$ and $p=0.64,U_1=2/3U_0$ would give the same $\braket{U_i}$ and $h^2$.

We shall express all energy scales in units of $\Delta_0$ - the superconductor order parameter in the clean limit ($L\rightarrow\infty, p=1$) in presenting our results. 
 The finiteness of the grain size $\mathcal{V}\sim L^d$ is reflected in our calculation by introducing a low-energy cutoff $\epsilon_c$ when we replace the sum over $k$ by an integral over single-particle energy $\xi_k$ in the mean field equations, i.e., $\frac{1}{\mathcal{V}}\sum_k\rightarrow N(0)\int^{\omega_D}_{\epsilon_c}d\xi$, where $\epsilon_c \sim 1/(N(0)L^d)$ is the energy-level spacing in a grain of size $L^d$. More details are presented in Appendix \ref{MFE}.

\subsection{Mean-field parameters}
 
\begin{figure}[htbp]
    \centering
    \begin{subfigure}[b]{0.8\columnwidth}
        \includegraphics[width=\linewidth]{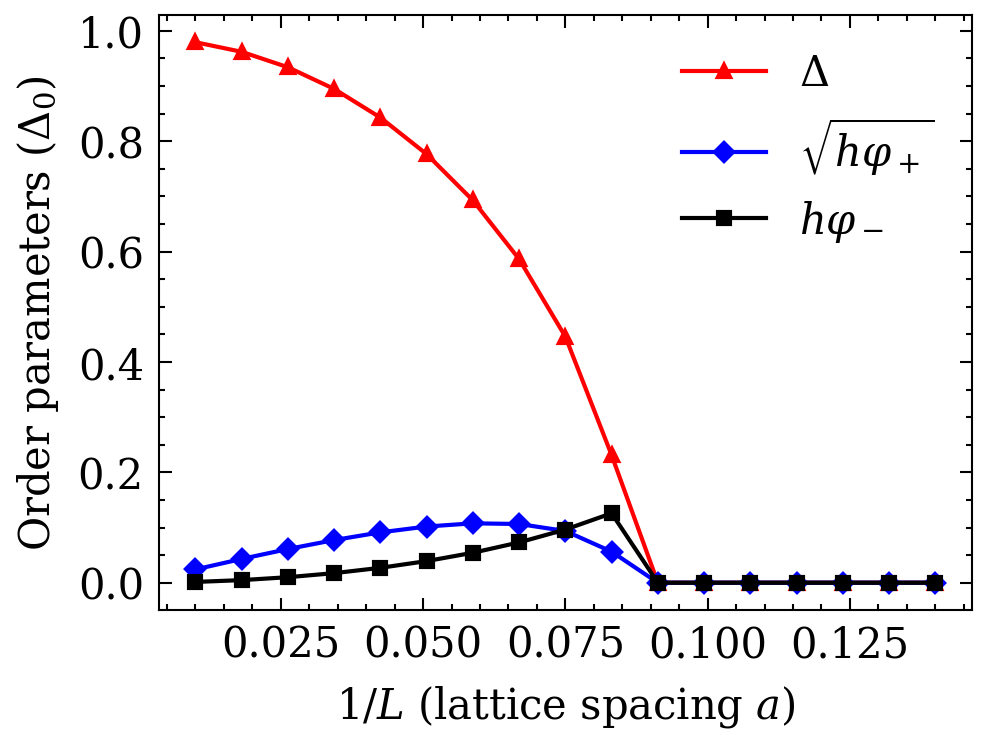}
        \caption{}
        \label{pd:a}
    \end{subfigure}
    \hfill 
    \begin{subfigure}[b]{0.8\columnwidth}
        \includegraphics[width=\linewidth]{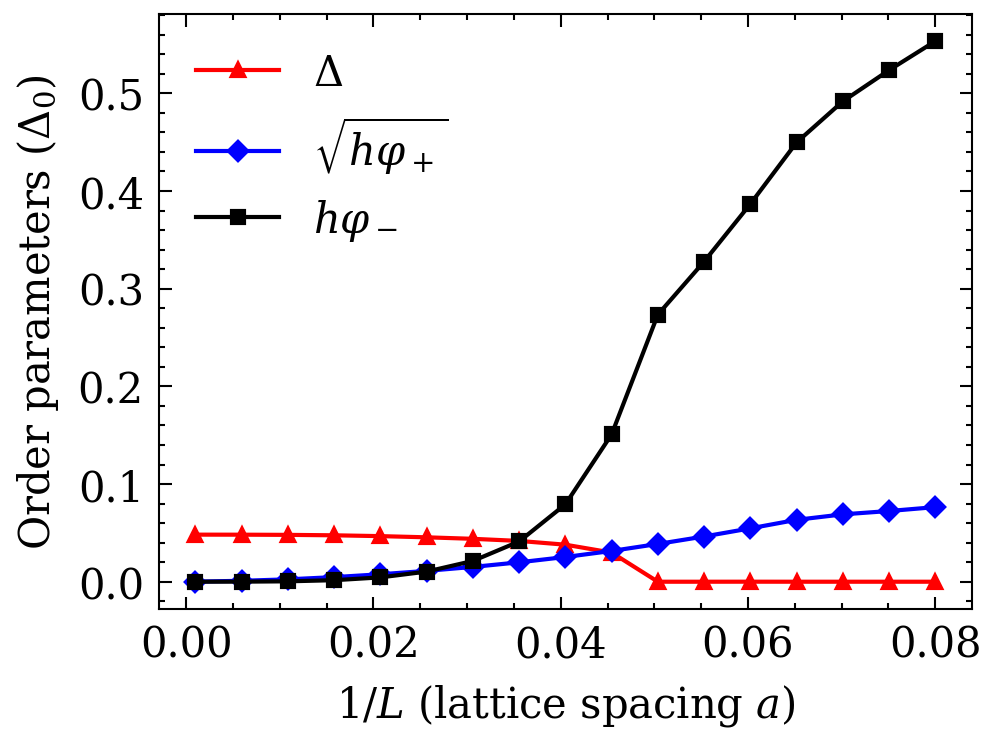}
        \caption{}
        \label{pd:b}
    \end{subfigure}
    \caption{The superconductor order parameter $\Delta$, Cooper-pair amplitude parameter $\sqrt{h\varphi_+}$ and CPG order parameter $h\varphi_-$ as a function of $1/L$ where $L$ is expressed in unit of lattice constant $a$. (a) $U_i=-U_0$, $p=1$, there is no glass state observed. (b) $d_I=15a$, $p=0.55, U_1=U_0$, the superconductor to CPG transition at $1/L_c \sim 0.05$ is clear from the figure. }    \label{pd}
\end{figure}

The self-consistently determined mean-field parameters $\Delta$, $h\varphi_-$ and $\sqrt{h\varphi_+}$ are presented in Fig.\ref{pd}. In the first case with pure attractive interaction we find that superconductivity persists until $L\sim 11$ lattice spacings, where $\Delta$ and $h\varphi_{\pm}$ all vanish, corresponds to the limit when the energy level spacing $\epsilon_c$ is comparable with the clean superconducting gap $\Delta_0$. There is no Cooper-pair glass (CPG) state found, in agreement with the expectation that existence of frustrated interaction is a necessary condition for the appearance of the CPG state. When Josephson coupling between grains are included, a Ginsburg-Landau type mean-field theory predicts that the superconductor (and other) order parameter will not vanish at $\epsilon_c\sim\Delta_0$, but will continue to exist until a much smaller grain size. The Josephson coupling will not lead to CPG state as it is not frustrated. We shall restrict ourselves to $L^d>1/(N(0)\Delta_0)$ in our following discussions.
 
 On the other hand, in the second case where regions with different signs of interaction exist, we find that the superconducting order parameter $\Delta(L^{-1}\rightarrow0)$ becomes much smaller than $\Delta_0$ because the presence of repulsive interaction regions lead to much smaller absolute value of $\braket{U_i}$. $\sqrt{h\varphi_+}$ and $h\varphi_-$ increase with disorder and there is a zero-temperature superconductor to CPG phase transition at a critical disorder strength $L_c \sim 20a $ where $\Delta$ vanishes. Increasing $h^2$ or decreasing magnitude of $\braket{U_i}$ facilitate the emergence of the CPG state. In the superconductor state, both $\Delta$ and $h\varphi_-$  are nonzero whereas $\Delta$ is zero in the CPG state. This is similar to classical spin glass models where the net magnetization becomes zero in the spin-glass state while the glass order parameter remains finite in both ferromagnetic and spin-glass states\cite{sherrington1975solvable, mezard1987spin, nishimori2001statistical,binder1986spin}. We note that our mean-field result becomes unreliable for $L^{-1}\ge0.067\sim d_I^{-1}$.

The mean-field equations can be analyzed analytically at small disorder. In the clean limit
 $h \to 0$ or $1/L \to 0$, we obtain $\varphi_+=\varphi_-=0$ and we recover the well-known BCS mean-field equation $1=\braket{U_i} N(0) \ln \frac{2 \omega_D}{\Delta(0)}$ where $\Delta(0)=\Delta(1/L \to 0)$. 
 
 For small disorder $1/L \ll 1$, we can expand the self-consistent equation in powers of $1/L$ and obtain to leading order in $1/L$,
  $\varphi_+ \sim \frac{N(0)\bar{U} \pi \Delta(0)}{8L}$, $\varphi_-\sim \frac{N(0)\bar{U}}{2L}(\ln{\frac{2\omega_D}{\Delta(0)}}-\frac{1}{2})$  and 
\begin{equation*}
\begin{aligned}
&~~~~~\Delta(1/L) \\
&\sim \Delta(0) + \frac{\pi N(0)^2 \bar{U}^3}{16 L^2}\left(\frac{1}{N(0) \bar{U}} - \frac{3}{2}\right)- 2\epsilon_c \\
&\sim \Delta(0) + \frac{\pi N(0)^2 \bar{U}^3}{16 L^2} \left(\frac{1}{N(0) \bar{U}} - \frac{3}{2} - \frac{32}{\pi N(0)^3 \bar{U}^3} \right).
\end{aligned}
\end{equation*}
 Notice that to order $L^{-2}$, the appearance of regions with different signs of interaction affects only $\bar{U}$. $\Delta(1/L)$ {\em decreases} as a function of $1/L$ at weak disorder, indicating that the multi-fractal effects that appear at weak disorder\cite{burmistrov2012enhancement, evers2008, stosiek2021multifractal} in infinite systems is absent in our study where we apply the replica trick to a single grain of size $L^d$.  

We can also perform a small $\bar{U}$ expansion at a fixed, finite value of $h$. In this case,  we can show that $\Delta=0$ (superconductivity destroyed) if $\sqrt{h\varphi_+} > \Delta_0(1+\bar{U}N(0)/c(h))$ where  $c(h)$ is a real number function of $h$ of order O(1). We also find that $\sqrt{h\varphi_+}\sim h^2N(0) +${\em logarithmic corrections} and the CPG state occurs only when \begin{equation}
    \label{criteria}
(\braket{U_i}^2+\frac{\sigma_U^2D_I^d}{L^d}) N(0)^2\geq L^d N(0)\Delta(0).
\end{equation}
In the weak coupling limit $\bar{U} N(0) \ll 1$ the inequality means that the glass state exists when $\sigma_U$ is large enough, $d_I/L\sim O(1)$ and $\Delta(0)$ is small. In particular, the CPS always exist when $\braket{U_i}=0$ and $\sigma_U\neq0$. This is possible only when regions with different signs of interaction exist. Numerically, we find that with the parameters we choose, the glass state exists when $L \lesssim 20a$ for $N(0)\bar{U}=0.05$, in qualitative agreement with the above criteria.

In the following we shall assume that the mean-field parameters computed for grains can be employed to study the properties of the macroscopic systems. The validity of the assumption will be discussed below.


\subsection{Single Particle Density of states}
\label{IIIB}
For a given configuration of disorder, the imaginary-time one-electron Green function is given in BCS theory by (see Appendix \ref{ME}, Eq.(A6))
\begin{equation}
    \mathcal{G}(k,i\omega_n)=\frac{|u_k|^2}{i\omega_n-E_k+U_{kk}/2}+\frac{|v_k|^2}{i\omega_n+E_k-U_{kk}/2}
\end{equation}
where the single-particle excitation energy is shifted from the BCS result for clean superconductors by $U_{kk}/2$ for localized single-particle wavefunctions. The single-particle density of states is given by the imaginary part of the retarded Green function 
\begin{equation}
\begin{aligned}
   &\quad D(\omega)  \\
     &= -\frac{1}{\pi}\frac{1}{V} \sum_k \text{Im}G^R(k,\omega) \\
    &= \frac{1}{V}\sum_k \left(|u_k|^2\delta(\omega-E_k+\frac{U_{kk}}{2})+|v_k|^2\delta(\omega+E_k-\frac{U_{kk}}{2})\right)
    \label{single dos}
\end{aligned}
\end{equation}
 We also obtain using Eq.\ (\ref{ss}), 
\begin{equation}
    |u(v)_k|^2=\frac{1}{2}(1+(-)s^z_k);  \quad \quad   E_k=\xi_k/s_k^z,
\label{uv}
\end{equation}
i.e., $D(\omega)$ is a function of $\{ s^z_k\}$. Therefore, the disorder-average over $P(U_{kp})$ can be obtained by computing $\braket{\braket{D(\omega)}}_d$ using Eq.\ (\ref{observable})(see Appendix \ref{PO} for details). We expect that the mean-field parameters computed for dirty superconductor grains can be used to compute $\braket{\braket{D(\omega)}}_d$ for macroscopic system as long as we are far away from the critical region where the order parameter $\Delta(L^{-1})$ vanishes. The Josephson coupling between grains is not going to modify the mean-field order parameters strongly as it is a local property of the system.

\begin{figure}[htbp]
    \centering
    \begin{subfigure}[b]{0.8\columnwidth}
        \includegraphics[width=\linewidth]{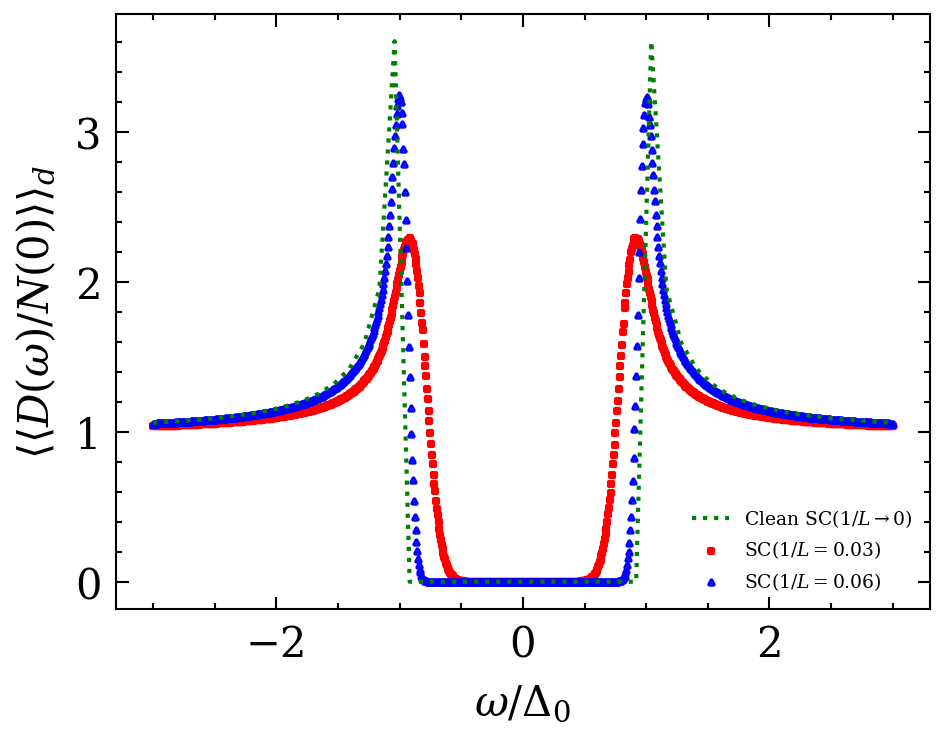}
        \caption{}
        \label{dos:a}
    \end{subfigure}
    \hfill 
    \begin{subfigure}[b]{0.8\columnwidth}
        \includegraphics[width=\linewidth]{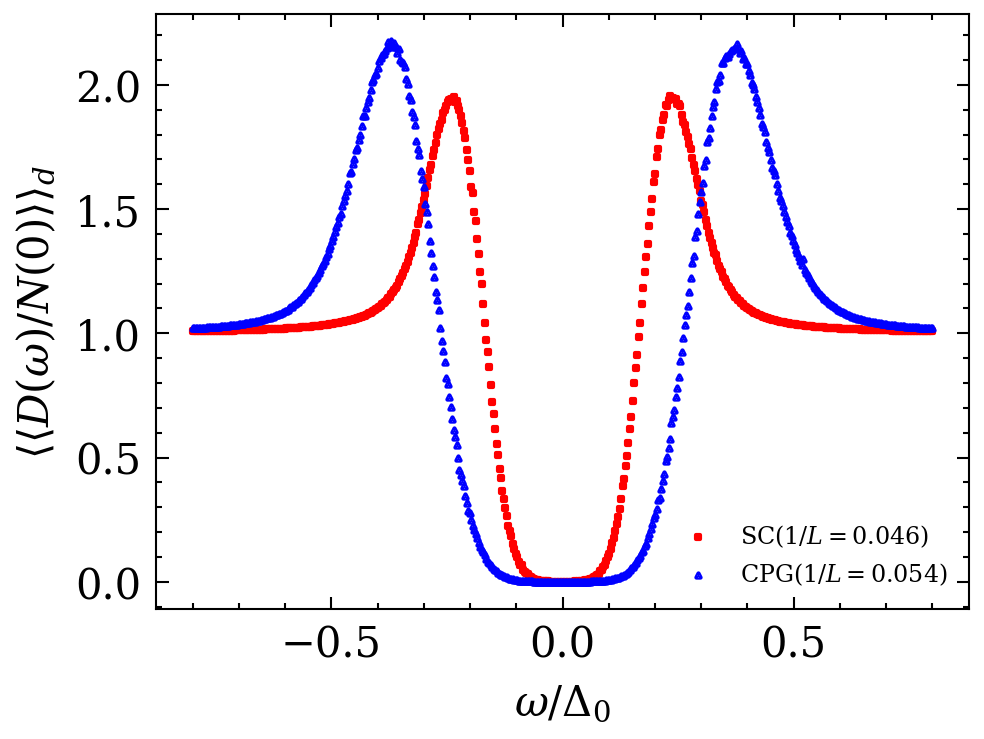}
        \caption{}
        \label{dos:b}
    \end{subfigure}
    \caption{The tunneling single-particle density of states are plotted for (a) $p=1$, $U_i=-U_0$, the spectral gap of the superconductor state remains rather robust to disorder. (b) $p=0.55, U_1=U_0$, the spectral gap is much reduced and smeared out. There is no abrupt change across the superconductor-CPG transition. We note that the coherence peak is smoothed out but remains in all cases.}
    \label{dosplot}
\end{figure}

When fluctuations in $U_{kk}$ is included, we also have to average over $U_{kk}$ and
\begin{equation}
    \braket{\braket{D(\omega)}}_d \rightarrow \int dU_{kk} P(U_{kk})\braket{\braket{D(\omega)}}_d 
\end{equation}
where $P(U_{kk})$ is the probability distribution of $U_{kk}$. We assume $P(U_{kk})=\frac{1}{\sqrt{2\pi \sigma_{k}^2}} e^{-\frac{(U_{kk}-U_{km})^2}{2\sigma_k}}$ in our numerical calculation, with $U_{km}=\bar{U}$ and $\sigma_k^2=\sigma^2$ (see Appendix A). The resulting Density of States (DOS) for $p=1$, $U_i=-U_0$ and $p=0.55$, $U_1=U_0$ are presented in Fig.\ref{dosplot} for two different values of $L^{-1}$. In Fig.\ref{dos:a} for $p=1$, we see that the spectral gap is slightly broadened and smeared out between $L^{-1}=0.03$ and $L^{-1}=0.06$, but remains rather robust towards disorder, although the superconductor order parameter $\Delta$ has already reduced by $\sim 50\%$, in qualitative agreement with previous numerical studies\cite{bouadim2011single, ghosal2001inhomogeneous}. 

In contrast, for $p=0.55$ where regions with different signs of interaction exist and $\Delta(0)$ is strongly reduced ($N(0)\bar{U}=0.05$), we find that (see Fig.\ref{dos:b}) the spectral gap is much reduced and is broadened and smeared out by disorder. There is no abrupt change in $D(\omega)$ across the transition from superconductor to CPG state. We note that the coherence peak remains in all the cases we study indicating that these states are all superconducting as we shall demonstrate in the following.

\subsection{Superfluid density}
We next compute the superfluid density or phase stiffness $\rho_s$ in mean-field theory. An approximate BCS mean-field current response (for macroscopic system) has been derived by De Gennes\cite{de2018superconductivity} where the superfluid density is expressed as 
\begin{subequations}
\label{rho}
\begin{equation}
\rho_s=\frac{2}{ \pi c} \int {d\xi} \int {d\xi'} L\left(\xi, \xi^{\prime}\right) \operatorname{Re} \sigma\left(\xi-\xi^{\prime}\right)-\frac{n e^2}{m c}
\end{equation}
where
\begin{equation}
\label{L}
L\left(\xi, \xi^{\prime}\right)=\frac{1}{2}\frac{EE'-\Delta^*_{\xi}\Delta_{\xi'}-\xi\xi'}{EE'(E+E')}
\end{equation}
\end{subequations}
with $E=\xi/s_{\xi}^z$. We have re-labeled the states by $k \to \xi_{k}=\xi, k' \to \xi_{k'}=\xi'$, etc. in writing down Eq.\ (\ref{rho}). $\operatorname{Re} \sigma\left(\xi-\xi^{\prime}\right)$ is the real part of the conductivity (averaged by disorder) in the normal state and $L(\xi,\xi')$ encodes the paramagnetic current response in the BCS mean field theory. More details of the derivation of Eq.\ (\ref{rho}) is given in Appendix \ref{PO}.  We caution that unlike the single-particle density of states, the superfluid density is a global property of the superconductor and depends strongly on the Josephson coupling between grains which is absent in our calculation. Therefore, our result should be considered to provide an order of magnitude estimate of $\rho_s$ only.   

\begin{figure}
\centering
\includegraphics[width=\columnwidth]{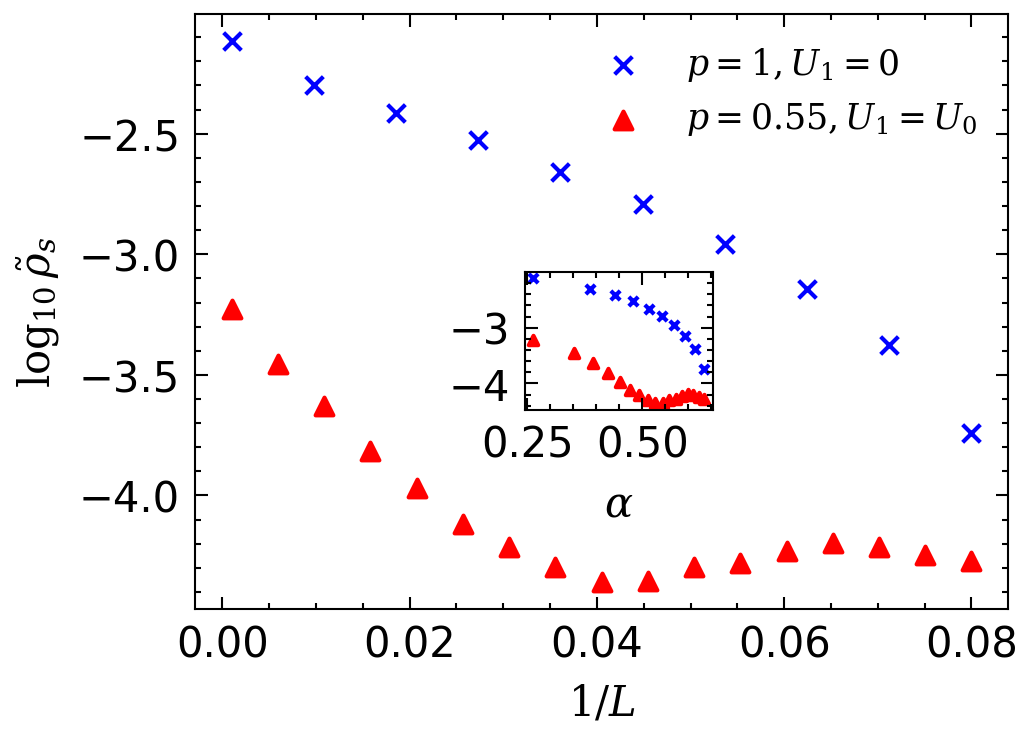}
\caption{The dimensionless superfluid density($\tilde{\rho}_s$) as a function of $1/L$ for both $p=1, U_i=-U_0$ and $p=0.55, U_1=U_0$. The rapid decrease in $\tilde{\rho}_s$ as function of disorder is clear. We observe that the superfluid density drops more rapidly in the $p=0.55$ case. The same result is plotted in the inset where we express the strength of disorder in units of normal state Drude resistance $\alpha = R/R_Q$, where $R_Q=h/e^2$ is the quantum resistance. We choose  $v_F/a=10^{16} s^{-1}$ and $k_Fa=1.5$ in our numerical calculation. With these parameters the clean superconductor coherence length $\sim \hbar v_F/\Delta_0$ is about $600a$.}
\label{sf}
\end{figure}

Similar to the density of states, the parameters $E=\xi/s^z_{\xi}$ and $\Delta_{\xi}=\frac{1}{2}(s^x_{\xi}+is^y_{\xi})$ are functions of $\vec{s}_{\xi}$ and we can perform the disorder-average of $\rho_s$ over $\vec{s}_{\xi}, \vec{s}_{\xi'}$ using Eq.\ (\ref{observable}) within the replica symmetric theory. We obtain
\begin{equation}
\label{dintegral}
 \quad \left\langle\braket{\rho_s}\right\rangle_d \\
=\frac{2}{ \pi c} \int \frac{d \bs{y}}{2 \pi} \frac{d \bs{y}^{\prime}}{2 \pi} e^{-\frac{\bs{y}^2+\bs{y'}^2}{2}} \braket{\rho_s \left(\bs{y}, \bs{y}^{\prime}\right)} \\
\end{equation}
where $\braket{\rho_s \left(\bs{y}, \bs{y}^{\prime}\right)}$ is obtained from Eq.(\ref{rho}) with $E(E')$ and $\Delta_\xi(\Delta_{\xi'})$ depending implicitly on $\bs{y}(\bs{y'})$ through $\vec{s}_{\xi}(\vec{s}_{\xi'})$. Note that we have to integrate over two independent random fields $\bs{y}$ and $\bs{y'}$ in the present case. The integrals are evaluated using numerical Monte Carlo method. Further details of the calculation are given in Appendix \ref{PO}.

We have computed the dimensionless superfluid density $\braket{\braket{\tilde{\rho}_s}}_d=\left\langle\braket{\rho_s}\right\rangle_d/(\frac{n e^2}{m c})$ numerically for different disorder strengths $1/L$ for both $p=1, U_i=-U_0$ and $p=0.55, U_1=U_0$. The results are shown in Fig.\ref{sf} in logarithmic scale. We note that the superfluid density decreases rapidly with increasing disorder in both cases with the superfluid density in the second case being much weaker because of smaller $\braket{U_i}$. The superfluid density remains nonzero in the CPG state, in agreement with numerical Monte Carlo calculation result for the random quantum rotor model\cite{granato2020disorder} and a peak appears around the region when $\Delta(L)$ vanishes and the CPG state emerges. 

For our choice of parameters, we find that for the first case with pure attractive interaction, $\braket{\braket{\tilde{\rho}_s}}_d \sim 3.7\Delta_0 \tau$ at weak disorder $k_Fl \gg 1$, (but with $\Delta_0\tau \ll 1)$ and drops more rapidly when the dimensionless Drude resistance $\alpha\geq 0.5$ (see inset), corresponding to the region $N(0)\Delta_0L^d\sim1$ where Anderson theorem breaks down, in agreement with previous results of Ma and Lee\cite{ma1985localized}. 

The fact that $\braket{\braket{\tilde{\rho}_s}}_d$  remains finite in the Cooper-pair glass state is not surprising as the superfluid density measures phase stiffness but not the order parameter $\Delta$ (See, for example Ref.\cite{ma1985localized}). 
An order of magnitude estimation shows that  $\braket{\braket{\tilde{\rho}_s}}_d$ in the CPG state is of order
\begin{equation}
\braket{\braket{\tilde{\rho}_s}}_d \sim \frac{l}{L}\left(\frac{h \varphi_{-}}{\omega_0}\right)^3 
\end{equation}
where $l$ is the scattering length and $\omega_0\sim$ energy scale below which the single particle states are localized.


\subsection{Cooper-pair glass state}
Although $\braket{\braket{\tilde{\rho}_s}}_d\neq0$ in both dirty superconductor state and Cooper-pair glass state, the Cooper-pair glass state differs from the usual dirty superconducting state. Theoretically, the difference can be seen most easily in the large-time behaviour of the pair-correlation function $\braket{P_{ij}(t)}_d=\braket{\braket{\Delta^\dagger_i(t)\Delta_j(0)}}_d$, where $\Delta_i^\dagger(t)=c^\dagger_{i\uparrow}(t)c^\dagger_{i\downarrow}(t)$ and $\Delta_i(t)=c_{i\downarrow}(t)c_{i\uparrow}(t)$ are the Cooper-pair creation and annihilation operators on site $i$, respectively.

The low energy behaviour of $P_{ij}(t\rightarrow\infty)$ is dominated by mean-field amplitudes where $\braket{P_{ij}(\infty)}_d\rightarrow\braket{\braket{\Delta^\dagger_i}\braket{\Delta_j}}_d$,
with
\[ \braket{\Delta_j}=\sum_{k,p}\braket{c_{k\downarrow}c_{p\uparrow}}\phi_{kj}\phi_{pj}=\sum_{k}\braket{(s^{x}_{k}+is^y_k)}|\phi_{kj}|^2. \]

We shall consider two limits in the following: (i)$|\vec{r}_i-\vec{r}_j|\rightarrow\infty$, and (ii)$\vec{r}_i=\vec{r}_j$. In the first case, the wavefunctions $|\phi_{ki}|^2$ and $|\phi_{pj}|^2$ are uncorrelated for localized states, and we obtain
\begin{subequations}
\begin{equation}
\braket{P_{\infty}(\infty)}_d\rightarrow\braket{\braket{\Delta^\dagger_i}}_d\braket{\braket{\Delta_j}}_d\sim(\frac{2\Delta}{\bar{U}})^2.
\end{equation}
which is nonzero only in the superconducting state. In the second case,
\begin{eqnarray}
\braket{P_0(\infty)}_d & = & \braket{\braket{\Delta^\dagger_i}\braket{\Delta_i}}_d \\ \nonumber
& \sim & (\frac{2\Delta}{\bar{U}})^2+\sum_k\braket{(s^x_k-is^y_k)(s^x_k+is^y_k)}_d|\phi_{ki}|^4  \\ \nonumber
& \sim & (\frac{2\Delta}{\bar{U}})^2+\frac{h\varphi_+}{h^22^d(L)^d}
\end{eqnarray}
\end{subequations}
where the second term is nonzero in the CPG state where $\Delta=0$ but $h\varphi_+\neq0$. Notice that $h\varphi_+\sim h^2N(0)\Delta_0$ when $h\rightarrow 0$ and the second term vanishes in the limit $L\rightarrow\infty$, i.e. when single-particle states become extended.

Experimentally, because of random configurations $\Delta_k$'s, it is expected that the CPG state will show strong hysteresis behaviour under external magnetic field. However, magnetic field may induce vortex glass states in dirty superconductors\cite{murray1990hexatic, sanchez2019vortexglass} and does not provide a sharp distinction between the vortex glass and CPG states.

A sharp distinction between the superconducting state and CPG state occurs when we consider Josephson effect between our targeted state and a normal s-wave superconductor in the absence of magnetic field. In usual tunneling between superconductors, the Josephson effect between two superconductors is proportional to the product of the anomalous Green's functions of the two superconductors (see for example, Ref.\cite{mahan2000many}), and is equal to zero when one of the superconductor is in the CPG state where $\sum_k\braket{\braket{F(k,i\omega_n)}}_d=0$ when $\Delta=0$. Physically, we expect that different spatial regions in the CPG state will be dominated by $\Delta_k$'s with different (and random) phases, resulting in vanishing average Josephson coupling between the CPG state and a normal s-wave superconductor. 

However, a higher order effect not captured by lowest-order tunneling theory exists as the coupling between the normal s-wave superconductor with order parameter $\Delta_n$ will induce a nonzero average superconductor order parameter $\Delta\sim\alpha(h)\Delta_n$ in the CPG state. As a result $\sum_k \braket{\braket{F(k,i\omega_n)}}_d\sim \alpha(h)\Delta_n$ becomes nonzero and a nonlinear Josepshon effect with Josephson current $j_s\sim\sqrt{\Delta_n\Delta}\sim\sqrt{\alpha(h)}\Delta_n$ will occur. The existence of this nonlinear Josephson effect distinguishes between the dirty superconductor state and CPG state.
The coefficient $\alpha(h)$ can be estimated by coupling our dirty superconductor to an external constant pairing field $\Delta_n$. After some straightforward algebra, we find that the effective pairing field $\bs{B}_k$ we introduced in the discussion after Eq.\ (\ref{observable}) is replaced by
\[\bs{B}_k\rightarrow\sqrt{\frac{h \varphi_{+}}{2}} \bs{y_k}+(\Delta+\Delta_n)\bs{\hat{x}}.  \]
The coefficient $\alpha(h)$ can be obtained by solving the corresponding mean-field equations in the CPG state.


\section{Discussion}
\label{disc}  
In this paper we develop a replica-BCS mean-field theory for dirty superconductors in zero magnetic field and apply it to study two dimensional superconductor grains of size $\sim L^2$, assuming that the macroscopic system is made up of weakly coupled grains. Besides dirty superconductors with pure attractive interaction, we also consider dirty superconductors where regions of attractive and repulsive interactions both exist in our study. 
To implement the replica-trick we made two approximations in our approach: we have employed the BCS mean-field theory instead of the more general Bogoliubov-de Gennes mean-field theory in describing superconductors\cite{de2018superconductivity, ghosal2001inhomogeneous}. With this approximation the effects of disorder are all encoded in the interaction matrix elements $U_{pk}$'s. We then made a very crude assumption about the distribution of matrix elements $U_{pk}$. In the case where attractive and repulsive interaction regions both exist, we made an assumption that the localized electronic states are insensitive to the boundary between these regions in our calculation. The goal of our paper is to see whether the known properties of dirty superconductors can be obtained qualitatively with these crude approximations for the case with pure attractive interaction and to see what the theory predicts when regions with attractive and repulsive interaction both exist.

  Our approximation limits ourselves to the cases when the relevant single-particle electronic states are localized, and we consider the replica-symmetric solution to the resulting mean-field equations. We find that the solution is stable against replica symmetry-breaking terms at zero temperature, in agreement with previous results on the disordered quantum rotor model\cite{ye1993solvable,read1995landau}. This is in contrast to temperature $T\neq0$ situations in which replica symmetry breaking terms are found to be important\cite{mezard1987spin,nishimori2001statistical,binder1986spin}.
  
 
 Our results for single-particle spectral gap and superfluid density are in qualitative agreement with results obtained from numerical QMC and simulation of BdG equation\cite{bouadim2011single, ghosal2001inhomogeneous} for dirty superconductors with pure attractive interaction. However the superconductor order parameter $\Delta(L)$ is found to decrease with disorder at weak disorder, suggesting that multi-fractal effect\cite{burmistrov2012enhancement, evers2008, stosiek2021multifractal} is not included in our study. This is not surprising in view of the very crude approximations made in our theory.
 
 When applying our method to dirty superconductors where regions of attractive and repulsive interactions both exist, a Cooper-pair-glass (CPG) phase is found when the mean interaction $\braket{U_i}$ is weak enough, the variance $\sigma_U^2$ is large enough and $d_I/L\sim O(1)$. The single-particle spectral gap is reduced and smeared out much more rapidly when compared with the case with pure attractive interaction and the superfluid density reduced much more rapidly but remains finite in the CPG phase. 


The regions where the superfluid density (or phase-stiffness) becomes very weak are of interests as these are the regions where superconductivity can be destroyed easily by quantum fluctuations and the anomalous metal state may emerge. In the case of dirty superconductors with pure attractive interaction, this happens when the system satisfies $\epsilon_c \sim 1/(N(0)L^d)\Delta_0$, which is the regime where the Anderson theorem breaks down. In this case, the superfluidity is expected to be destroyed by the Coulomb blockade effect and the system becomes an insulator in a fully quantum mechanical treatment of the problem\cite{mpafisher1986,fisher_1990,fazio1991charge, Sacepa2020quantum, beloborodov2007granular}. Metallic behavior is not expected in this case as the singlar-particle spectrum of the system is still fully gaped.

  The case of dirty superconductors with both attractive and repulsive interaction regions is more interesting as the superfluid density is in general much weaker and the single-particle energy gap is much reduced and smeared out, in particular when the system approaches the CPG state where large regions of attractive and repulsive regions coexist. Notice that this may happen for an arbitrarily large localization length $L$ when $\Delta(0)\rightarrow 0$ according to Eq. (\ref{criteria}), suggesting that this may be a favorable candidate as a parent state for the anomalous metal state. However, we caution that the electronic states are more likely to localize within regions of fixed interaction when $d_I\sim L$. In this case, a more realistic starting point is to consider a model of mixed metal-superconductor grains. This aligns with earlier proposals that a model of superconductor grains immersed in metallic sea is a favorable parent model for the anomalous metal state\cite{kapitulnik2019colloquium, Sacepa2020quantum,  fl, sz, sk}.

 It is important to understand the dynamics of the dirty superconductor and CPG states at energies below the single-particle spectral gap because it is directly related to whether/how these states can be destroyed by quantum fluctuations and whether the anomalous metal states may emerge as a result. 
The collective excitation spectrum of dirty superconductors can be studied by generalizing the present mean-field theory to include dynamic (Gaussian) fluctuations in replica theory. We shall carry out the study in a future paper. 

Lastly, we comment that with suitable modifications, our replica approach here can be generalized to study mean-field theories of other quantum systems with broken symmetry and disorder. 


\begin{acknowledgments}
The project is supported by the School of Science, the Hong Kong University of Science and Technology.
\end{acknowledgments}


\begin{thebibliography}{99}
\bibitem{kapitulnik2019colloquium} Aharon Kapitulnik, Steven A. Kivelson, and Boris Spivak, Colloquium: Anomalous metals: Failed superconductors, \href{https://doi.org/10.1103/RevModPhys.91.011002}{Rev. Mod. Phys. \textbf{91}, 011002 (2019)}.

\bibitem{Sacepa2020quantum}
B. Sacépé, M. Feigel’man, and T. M. Klapwijk, 
Quantum breakdown of superconductivity in low-dimensional materials, 
\href{https://doi.org/10.1038/s41567-020-0905-x}{Nature Physics \textbf{16}, 734--746 (2020)}.

\bibitem{wang2023quantum}
Ziqiao Wang, Yi Liu, Chengcheng Ji, and Jian Wang,
Quantum phase transitions in two-dimensional superconductors: a review on recent experimental progress,
\href{https://doi.org/10.1088/1361-6633/ad14f3}{Rep. Prog. Phys. \textbf{87}, 014502 (2024)}.

\bibitem{bouadim2011single}
Karim Bouadim, Yen Lee Loh, Mohit Randeria, and Nandini Trivedi,
Single- and two-particle energy gaps across the disorder-driven superconductor–insulator transition,
\href{https://doi.org/10.1038/nphys2037}{Nat. Phys. \textbf{7}, 884-889 (2011)}.


\bibitem{read1995landau}
N. Read, Subir Sachdev, and J. Ye,
Landau theory of quantum spin glasses of rotors and Ising spins,
\href{https://doi.org/10.1103/PhysRevB.52.384}{Phys. Rev. B \textbf{52}, 384 (1995)}.

\bibitem{2023intrinsic}
Yat Fan Lau and Tai Kai Ng,
Intrinsic instabilities in Fermi glasses,
\href{https://doi.org/10.1103/PhysRevB.110.075153}{Phys. Rev. B \textbf{110}, 075153 (2024)}.

\bibitem{ghosal2001inhomogeneous}
A. Ghosal, M. Randeria, and N. Trivedi, 
Inhomogeneous pairing in highly disordered s-wave superconductors, 
\href{https://doi.org/10.1103/PhysRevB.65.014501}{Phys. Rev. B \textbf{65}, 014501 (2001)}; A. Ghosal, M. Randeria, and N. Trivedi, 
Role of Spatial Amplitude Fluctuations in Highly Disordered s-Wave Superconductors, 
\href{https://doi.org/10.1103/PhysRevLett.81.3940}{Phys. Rev. Lett. \textbf{81}, 3940 (1998)}

\bibitem{ioffe2010disorder}
L. B. Ioffe and Marc Mézard,
Disorder-Driven Quantum Phase Transitions in Superconductors and Magnets,
\href{https://doi.org/10.1103/PhysRevLett.105.037001}{Phys. Rev. Lett. \textbf{105}, 037001 (2010)}.

\bibitem{sknepnek2004order}
Rastko Sknepnek, Thomas Vojta, and Rajesh Narayanan,
Order-parameter symmetry and mode-coupling effects at dirty superconducting quantum phase transitions,
\href{https://doi.org/10.1103/PhysRevB.70.104514}{Phys. Rev. B \textbf{70}, 104514 (2004)}.

\bibitem{feigel2000keldysh} M. V. Feigel’man, A. I. Larkin, and M. A. Skvortsov, Keldysh action for disordered superconductors, \href{https://doi.org/10.1103/PhysRevB.61.12361}{Phys. Rev. B \textbf{61}, 12361 (2000)}.


\bibitem{phillips2003elusive}
Philip Phillips and Denis Dalidovich,
The Elusive Bose Metal,
\href{https://doi.org/10.1126/science.1088253}{Science \textbf{302}, 243-247 (2003)}.

\bibitem{ye1993solvable}
J. Ye, S. Sachdev, and N. Read,
Solvable spin glass of quantum rotors,
\href{https://doi.org/10.1103/PhysRevLett.70.4011}{Phys. Rev. Lett. \textbf{70}, 4011 (1993)}.


\bibitem{ma1985localized}
Michael Ma and Patrick A. Lee,
Localized superconductors,
\href{https://doi.org/10.1103/PhysRevB.32.5658}{Phys. Rev. B \textbf{32}, 5658 (1985)}.

\bibitem{kapitulnik1985anderson}
A. Kapitulnik and G. Kotliar,
Anderson Localization and the Theory of Dirty Superconductors,
\href{https://doi.org/10.1103/PhysRevLett.54.473}{Phys. Rev. Lett. \textbf{54}, 473 (1985)}.


\bibitem{granato2020disorder}
Enzo Granato,
Disorder-induced superconductor to insulator transition and finite phase stiffness in two-dimensional phase-glass models,
\href{https://doi.org/10.1103/PhysRevB.102.184503}{Phys. Rev. B \textbf{102}, 184503 (2020)}.

\bibitem{beloborodov2007granular} I. S. Beloborodov, A. V. Lopatin, V. M. Vinokur, and K. B. Efetov, Granular electronic systems, \href{https://doi.org/10.1103/RevModPhys.79.469}{Rev. Mod. Phys. \textbf{79}, 469 (2007)}.

\bibitem{anderson1959theory}
P. W. Anderson,
Theory of dirty superconductors,
\href{https://doi.org/10.1016/0022-3697(59)90036-8}{J. Phys. Chem. Solids \textbf{11}, 26-30 (1959)}.


\bibitem{sherrington1975solvable}
David Sherrington and Scott Kirkpatrick,
Solvable Model of a Spin-Glass,
\href{https://doi.org/10.1103/PhysRevLett.35.1792}{Phys. Rev. Lett. \textbf{35}, 1792 (1975)}.

\bibitem{fisher1989boson}
Matthew P. A. Fisher, Peter B. Weichman, Geoffrey Grinstein, and Daniel S. Fisher,
Boson localization and the superfluid-insulator transition,
\href{https://doi.org/10.1103/PhysRevB.40.546}{Phys. Rev. B \textbf{40}, 546 (1989)}.

\bibitem{weichman2008particle}
Peter B. Weichman and Ranjan Mukhopadhyay,
Particle-hole symmetry and the dirty boson problem,
\href{https://doi.org/10.1103/PhysRevB.77.214516}{Phys. Rev. B \textbf{77}, 214516 (2008)}.

\bibitem{mezard1987spin} M. Mézard, G. Parisi, and M. A. Virasoro, Spin glass theory and beyond: An Introduction to the Replica Method and Its Applications (World Scientific Publishing Company, Singapore, 1987), Vol. 9.

\bibitem{nishimori2001statistical}
Hidetoshi Nishimori,
Statistical physics of spin glasses and information processing: an introduction,
(Clarendon Press, Oxford, 2001). Number 111.

\bibitem{binder1986spin}
K. Binder and A. P. Young,
Spin glasses: Experimental facts, theoretical concepts, and open questions,
\href{https://doi.org/10.1103/RevModPhys.58.801}{Rev. Mod. Phys. \textbf{58}, 801 (1986)}.

\bibitem{mpafisher1986}
Matthew P. A. Fisher,
Quantum phase transitions in disordered two-dimensional superconductors,
\href{https://doi.org/10.1103/PhysRevLett.57.885}{Phys. Rev. Lett. \textbf{57}, 885 (1986)}.

\bibitem{fisher_1990}
Matthew P. A. Fisher, G. Grinstein, and S. M. Girvin,
Presence of quantum diffusion in two dimensions: Universal resistance at the superconductor-insulator transition,
\href{https://doi.org/10.1103/PhysRevLett.64.587}{Phys. Rev. Lett. \textbf{64}, 587 (1990)}.

\bibitem{fazio1991charge}
Rosario Fazio and Gerd Schön,
Charge and vortex dynamics in arrays of tunnel junctions,
\href{https://doi.org/10.1103/PhysRevB.43.5307}{Phys. Rev. B \textbf{43}, 5307 (1991)}.


\bibitem{burmistrov2012enhancement}
I. S. Burmistrov, I. V. Gornyi, and A. D. Mirlin,
Enhancement of the Critical Temperature of Superconductors by Anderson Localization,
\href{https://doi.org/10.1103/PhysRevLett.108.017002}{Phys. Rev. Lett. \textbf{108}, 017002 (2012)}.

\bibitem{evers2008}
Ferdinand Evers and Alexander D. Mirlin,
Anderson transitions,
\href{https://doi.org/10.1103/RevModPhys.80.1355}{Rev. Mod. Phys. \textbf{80}, 1355 (2008)}.

\bibitem{stosiek2021multifractal}
M. Stosiek, F. Evers, and I. S. Burmistrov,
Multifractal correlations of the local density of states in dirty superconducting films,
\href{https://doi.org/10.1103/PhysRevResearch.3.L042016}{Phys. Rev. Research \textbf{3}, L042016 (2021)}.

\bibitem{de2018superconductivity}
P. G. De Gennes,
Superconductivity of metals and alloys,
(CRC press, 2018).


\bibitem{murray1990hexatic}
C. A. Murray, P. L. Gammel, D. J. Bishop \textit{et al.},
Observation of a hexatic vortex glass in flux lattices of the high-$T_c$ superconductor Bi$_{2.1}$Sr$_{1.9}$Ca$_{0.9}$Cu$_2$O$_{8+\delta}$,
\href{https://doi.org/10.1103/PhysRevLett.64.2312}{Phys. Rev. Lett. \textbf{64}, 2312 (1990)}.






\bibitem{sanchez2019vortexglass}
J. Aragón Sánchez, R. Cortés Maldonado, N. R. Cejas Bolecek \textit{et al.},
Unveiling the vortex glass phase in the surface and volume of a type-II superconductor,
\href{https://doi.org/10.1038/s42005-019-0225-6}{Commun. Phys. \textbf{2}, 143 (2019)}.

\bibitem{mahan2000many}
G. D. Mahan,
Many-particle physics,
(Springer Science \& Business Media, 2013).

\bibitem{fl}
M. V. Feigel'man, A. I. Larkin, and M. A. Skvortsov,
Quantum superconductor--metal transition in a proximity array,
\href{https://doi.org/10.1070/1063-7869/44/10S/S22}{Phys.-Usp. \textbf{44}, 99 (2001)}.

\bibitem{sz}
B. Spivak, A. Zyuzin, and M. Hruska,
Quantum superconductor-metal transition,
\href{https://doi.org/10.1103/PhysRevB.64.132502}{Phys. Rev. B \textbf{64}, 132502 (2001)}.

\bibitem{sk}
B. Spivak, P. Oreto, and S. A. Kivelson,
Theory of quantum metal to superconductor transitions in highly conducting systems,
\href{https://doi.org/10.1103/PhysRevB.77.214523}{Phys. Rev. B \textbf{77}, 214523 (2008)}.






\bibitem{de1978stability}
J.R.L. de Almeida and David J. Thouless,
Stability of the Sherrington-Kirkpatrick solution of a spin glass model,
\href{https://doi.org/10.1088/0305-4470/11/5/028}{J. Phys. A: Math. Gen. \textbf{11}, 983 (1978)}.

\bibitem{vollhardt1980diagrammatic}
D. Vollhardt and P. Wölfle,
Diagrammatic, self-consistent treatment of the Anderson localization problem in $d \leq 2$ dimensions,
\href{https://doi.org/10.1103/PhysRevB.22.4666}{Phys. Rev. B \textbf{22}, 4666 (1980)}.
















\end{thebibliography}

\onecolumngrid 
\newpage

\appendix
\section{{Estimation of the interaction matrix elements}}
\label{ME}
Here we estimate the interacting matrix elements $U_{k p}$. We assume there exists both regions of attractive interaction($-U_0$) (with probability $p$) and repulsive interaction ($+U_1$) (with probability $1-p$) in the system. The size of region is $\sim d_I^d$. We assume that the electronic wavefunctions are insensitive to the boundaries between regions with different interactions and consider localized wavefunction $\left|\phi_k(\bs{x})\right|^2 \sim \frac{1}{L^d} e^{-\frac{\left|\bs{x}-\bs{X}_k\right|}{L}}$ with center of localization $\bs{X}_k$ randomly distributed in the system, $L$ is the localization length and $d$ is the spatial dimension. 

For $\bs{X}_k$ and $\bs{X}_p$ within a grain of size $L$ from each other, 
\begin{equation}
\begin{aligned}
\braket{U_{k p}} & \sim \sum_{i} \braket{U_{i}}\braket{\left|\phi_{ki}\right|^2\left|\phi_{pi}\right|^2} \\
&\approx \braket{U_i}(\frac{1}{L^{2d}})(L^d) = \frac{\braket{U_i}}{L^d}
\end{aligned}
\end{equation}
where $\braket{U_i}=-pU_0+(1-p)U_1$. The factor $1/L^{2d}$ comes from the normalization factor of localized wavefunctions and $(L^d)$ results from the sum over a volume of $L^d$ in which the overlap between two localized states are significant. We note that $U_{kp}\rightarrow0$ for $k$ and $p$ states separated by distance $\gg L$. 

Next we consider $\braket{U_{kp}U_{pk}}$, assuming $\bs{X}_k$ and $\bs{X}_p$ are within the same grain. In this case
\begin{equation}
\begin{aligned}
    \braket{U_{kp}U_{pk}} & \sim \sum_{i,j} \braket{U_iU_j}\braket{\left|\phi_{ki}\right|^2\left|\phi_{pi}\right|^2\left|\phi_{kj}\right|^2\left|\phi_{pj}\right|^2} \\
    & \sim   \sum_{ij}\left(\braket{U_i}^2+\sigma_U^2\theta(d_I-|\vec{x}_i-\vec{x}_j|\right))\times\left(\braket{|\phi_{ki}|^2|\phi_{pi}|^2|\phi_{kj}|^2|\phi_{pj}|^2}_c+\braket{\left|\phi_{ki}\right|^2\left|\phi_{pi}\right|^2}\braket{\left|\phi_{kj}\right|^2\left|\phi_{pj}\right|^2}  \right)  \\
    & \sim \braket{U_i}^2\left(\frac{\beta}{L^{2d}}+\frac{1}{L^{2d}}\right)+\sigma_U^2\sum_{i,j}\theta(d_I-|\vec{r}_i-\vec{r}_j|)\frac{1}{L^{4d}}  \\
    &\sim (\beta+1)\frac{\braket{U_i}^2}{L^{2d}}+\frac{\sigma_U^2 D_I^d}{L^{3d}},
\end{aligned}
\end{equation}
 where $\sigma_{U}=(U_0+U_i)\sqrt{p(1-p)}$.  \(\left\langle \left|\phi_{ki}\right|^2 \left|\phi_{pi}\right|^2 \left|\phi_{kj}\right|^2 \left|\phi_{pj}\right|^2 \right\rangle_c\) are correlated averages. We have approximated \(\left\langle \left|\phi_{ki}\right|^2 \left|\phi_{pi}\right|^2 \right\rangle \sim \frac{1}{L^{2d}}\) and \(\left\langle \left|\phi_{ki}\right|^2 \left|\phi_{pi}\right|^2 \left|\phi_{kj}\right|^2 \left|\phi_{pj}\right|^2 \right\rangle_c \sim \frac{\beta}{L^{4d}}\).
 $\beta$ is a coefficient of order $O(1)$. \( D_I = d_I \) for \( d_I \leq L \) and \( D_I = L \) for \( d_I > L \) as the whole grain will be in a region of fixed interaction in this case. Thus, the variance of $U_{kp}$ for arbitrary states $k,p$ within a grain with volume $L^d$ is given by
\begin{equation}
    \braket{U_{kp}U_{pk}}-  \braket{U_{kp}}^2 \sim \frac{\sigma_U^2D_I^d}{L^{3d}} +\beta \frac{\braket{U_i}^2}{L^{2d}}.
  \end{equation}
 Comparing with Eq.\ (\ref{Gaussian}) we obtain $h^2\sim \frac{\sigma_U^2D_I^d}{L^{2d}}+\frac{\braket{U_i}^2}{L^{2d}}$ where we set $\beta=1$ in the above expressions. We note that the above estimations are valid for $k=p$ also.
\\ \\
Next we consider the role of the $U_{kk}$ term in our theory. We note that it can be treated exactly in the single $k$ component of the BCS-Hamiltonian (see Eq.\ (\ref{mfd})) where
\begin{equation}
H_k=\sum_{\sigma}\xi_{k}c^\dagger_{k\sigma} c_{k\sigma} +\lambda_kc^\dagger_{k\uparrow}c^\dagger_{-k\downarrow}+\lambda_k^*c_{k\downarrow}c_{-k\uparrow}-U_{kk}n_{k\uparrow}n_{-k\downarrow}
\end{equation}
which can be diagonalized easily in the basis \{$\ket{k\uparrow -k\downarrow},\ket{0},\ket{k\uparrow},\ket{-k\downarrow}$\}. It is easy to see that the eigenstates are separated into even and odd fermion parity sectors \{$\ket{k\uparrow -k\downarrow},\ket{0}$\} and \{$\ket{k\uparrow},\ket{-k\downarrow}$\} and the eigenvalues of $H_k$ are given by
\begin{eqnarray}
    E_{\text{odd}} & = & \xi_k (\text{odd-fermion parity, doubly-degenerate}), \\ \nonumber
    E_{\text{even}} & = & \xi_k-U_{kk}/2\pm \sqrt{(\xi_k-U_{kk}/2)^2+|\lambda_k|^2} (\text{even-fermion parity}).
\end{eqnarray}
 Shifting $\xi_k-U_{kk}/2\rightarrow\xi_k$, it is easy to see that even-fermion parity sector and the BCS ground state is not affected by $U_{kk}$. $U_{kk}$ affects only the single-particle excitation energy $\xi_k+U_{kk}/2-(\xi_k-\sqrt{\xi_k^2+|\lambda_k|^2})=\frac{U_{kk}}{2}+\sqrt{\xi_k^2+|\lambda_k|^2}$, leading to the expression we used in calculating the single electron Green's function $\braket{\mathcal{G}(k,i\omega_n)}$ and the DOS $\braket{\braket{D(\omega)}}_d$.

\section{Derivation of mean-field self-consistent equations}
\label{MFE}
Here we provide some detailed derivation of the self-consistent mean field equations. We start with computing $\braket{Z^n}_d$ and then the disorder-averaged free energy by taking the $n \to 0$ limit. We then minimize the free energy with respect to the mean-field order parameters to obtain the self-consistent equations. For convenience we define $\bar{U}=|\braket{U_i}|$. We start with
 
\begin{equation}
\braket{Z^n}_d=\int \prod_{i=1, n} D S_i e^{-\beta \sum_{i=1}^n \sum_k \left( -\xi_k s_{k i}^{(z)}\right)} \int D\left[U_{k p}\right] P\left(\left[U_{k p}\right]\right) e^{\frac{\beta}{4} \sum_{k \neq p} U_{k p} \sum_{i=1}^n\left(s_{k i}^{(x)} s_{p i}^{(x)}+s_{k i}^{(y)} s_{p i}^{(y)}\right)}
\end{equation}
where $D\left[U_{k p}\right]=\prod_{k p} d U_{k p},  P\left(\left[U_{k p}\right]\right)=\prod_{k p} P\left(U_{k p}\right)$ and $P(U_{kp}) \sim \sqrt{\frac{\mathcal{V}}{2\pi h^2}} \exp{\left[\left({-\frac{\mathcal{V}(U_{kp}-\bar{U}/\mathcal{V})^2}{2h^2}}\right)\right]}$. Notice that the $U_{kk}$ terms are not involved in calculation of ground state properties as we explained in Appendix \ref{ME}. \\ \\
After integrating out $D\left[U_{k p}\right]$, we obtain
\begin{equation}
\braket{Z^n}_d=\int \prod_{i=1, n} D S_i \exp{\left[-\beta  \sum_k \left(\sum_{i=1}^n -\xi_k s_{k i}^{(z)}\right)+\sum_{k \neq p}\left[\frac{(h \beta)^2}{32\mathcal{V}}\left(\varphi_{k p}\right)^2+\frac{\beta}{4\mathcal{V}} \bar{U} \varphi_{k p}\right]\right]}
\end{equation}
where $\varphi_{k p}=\sum_{i=1}^n\left(s_{k i}^{(x)} s_{p i}^{(x)}+s_{k i}^{(y)} s_{p i}^{(y)}\right)$.
We can write
\begin{equation}
\begin{aligned}
\sum_{k \neq p}\left(\varphi_{k p}\right)^2
&=\sum_{i \neq j}\left(\left(\gamma_{i j}^{x x}\right)^2+\left(\gamma_{i j}^{y x}\right)^2+\left(\gamma_{i j}^{x y}\right)^2+\left(\gamma_{i j}^{y y}\right)^2\right)+\sum_{i \alpha} \kappa_i^\alpha \kappa_i^\alpha+2 \sum_i \kappa_i^0 \kappa_i^0
\end{aligned}
\end{equation}
where $\gamma_{i j}^{\alpha \beta}=\sum_k s_{k i}^{(\alpha)} s_{k j}^{(\beta)}, \kappa_i^\alpha=\sum_k s_{k i}^{(\alpha) 2}, \kappa_i^0=\sum_k s_{k i}^{(y)} s_{k i}^{(x)}, \alpha, \beta=x, y$ and \\
\begin{equation}
\label{partition_func}
    \braket{Z^n}_d=\int \prod_{i=1, n} D S_i e^{-\beta  \sum_k \left( \sum_{i=1}^n -\xi_k s_{k i}^{(z)} \right)+\frac{(h \beta)^2}{32\mathcal{V}} \sum_{i \neq j}\sum_{\alpha \beta}\left(\gamma_{i j}^{\alpha \beta}\right)^2+\frac{(h \beta)^2}{32\mathcal{V}} \sum_{i \alpha} (\kappa_i^\alpha)^2 +\frac{(h \beta)^2}{16\mathcal{V}} \sum_i (\kappa_i^0)^2+\frac{\beta}{4\mathcal{V}} \bar{U} \sum_i \tilde{S}_i^2}
\end{equation}
where $\tilde{S}_i = \sum_k(s_{k i}^{(x)} \bs{\hat{x}}+s_{k i}^{(y)} \hat{y})$. \\ \\
Next we introduce three Hubbard-Stratonovitch fields $q_{ij}^{\alpha \beta}, \chi_i^{\alpha}$ and $\chi^0_{i}$ to decouple the quadratic terms associated with $\gamma_{ij}^{\alpha\beta}, \kappa_i^{\alpha}$ and $\kappa_i^0$, respectively in Eq.(\ref{partition_func}). The resulting Gaussian integrals can be treated in the saddle-point approximation as the partition function represents a system with infinite-range interaction in $k$-space. For simplicity, we consider replica symmetric saddle point solutions without breaking rotational symmetry, i.e. we consider $q_{i j}^{\alpha \beta} = q \delta_{\alpha \beta},\quad \chi_i^\alpha = \chi \text {   and   }  \chi_i^0 = 0$ with $\bar{S}^\alpha_k = \sum_i s^\alpha_{ki}$.

The resulting partition function for $n$-replica is 
\begin{equation}
\begin{alignedat}{3}
    \braket{Z^n}_d=\int \prod_{i=1}^n &D[S_i] \exp{\left[-\beta  \sum_k \left( \sum_{i=1}^n -\xi_k s_{k i}^{(z)}) \right)+\frac{\beta}{4\mathcal{V}} \bar{U} \sum_i \tilde{S}_i^2\right]} 
     \\
    &\quad  \exp{\left[\frac{\beta hq}{4} \sum_{k \alpha} (\bar{S}^{\alpha}_{k})^2 -n^2\mathcal{V}q^2+\frac{\beta h}{4} \sum_{i k\alpha}  (s^{\alpha}_{ki})^2(\chi-q)-n\mathcal{V}(\chi^2-q^2)\right]}.
\end{alignedat}
\end{equation}

Next we introduce another two Gaussian integrals to decouple the remaining quadratic terms $\exp{\left[\frac{\beta U_0}{4\mathcal{V}}\tilde{S}_i^2\right]}$ and $\exp{\left[\frac{\beta h q_\alpha}{4}\left(\bar{S}_k^\alpha\right)^2\right]}$ through two Hubbard-Stratonovitch fields $\bs{b_i}$ and $\bs{y_{k}}$. We employ the saddle point approximation again to treat the $\bs{b_i}$ field which is equivalent to BCS mean-field theory but such an approximation is not applied to the $\bs{y_{k}}$ field as there is no justification, resulting in
\begin{equation}
\begin{aligned}
    \braket{Z^n}_d
    &=\int \prod_{k \alpha} \frac{\text{d} y_{k \alpha}}{\sqrt{2 \pi}} e^{-\frac{1}{2}\sum_{k\alpha}y_{k\alpha}^2} 
     \left(\prod_k \int D[S_k] e^{-\beta  \left( -\xi_k s_{k}^{(z)}-\sum_{\alpha} \frac{1}{4}h\varphi_{-}(s^{(\alpha)}_{k})^2 -\sum_{\alpha} \left(\sqrt{\frac{h\widetilde{q}}{2}}y_{k\alpha}+\Delta\delta_{\alpha x}\right)s^{(\alpha)}_{k} \right)} \right)^n \\
     & \quad \quad \exp{\left[-\frac{\beta n\mathcal{V} \Delta^2}{\bar{U}}-\frac{\beta^2n^2 \mathcal{V}}{2}\sum_{\alpha} \widetilde{q}^2-\beta n\mathcal{V}\sum_\alpha \varphi_{+} \varphi_{-}\right]}
\end{aligned}
\end{equation}
where we define $\sqrt{\frac{\beta U_0}{2}}b=\beta \Delta$, $q = \beta \widetilde{q}$, $q + \chi=2\beta \varphi_{+}$ and $\chi-q=\varphi_{-}$. We also assume $\bs{b_i}=\bs{b}=b\bs{\hat{x}}$ (we take $\bs{b}$ in $\bs{\hat{x}}$ direction without loss of generality). 

The disordered averaged mean free energy is given by
\begin{equation}
\begin{aligned}
\braket{F}_d&=-\lim _{n \rightarrow 0} \frac{\braket{Z^n}_d-1}{n \beta} \\
&= -\sum_k \int \frac{d \bs{y_{k}}}{2 \pi} \exp{\left[-\frac{\left(\bs{y_{k}}\right)^2}{2}\right]} \frac{1}{\beta} \ln \left(Z_{0 k}\right)+\frac{\mathcal{V}\Delta^2}{\bar{U}}+2\mathcal{V} \varphi_{+} \varphi_{-}
\end{aligned}
\end{equation}
where $d\bs{y_k}=\prod_\alpha d y_{k\alpha}$ and 
\begin{equation}
Z_{0 k}=\int D [S_k] e^{-\beta \left[ \left( -\xi_k s_{k}^{(z)}-\sum_{\alpha} \frac{1}{4}h\varphi_{-}(s^{(\alpha)}_{k})^2 -\sum_{\alpha} \left(\sqrt{\frac{h\widetilde{q}}{2}}y_{k\alpha}+\Delta\delta_{\alpha x}\right)s^{(\alpha)}_{k} \right) \right]}
\end{equation}
The self-consistent mean-field equations are obtained by minimizing the mean-field parameters with respect to $\braket{F}_d$. We obtain 
\begin{equation}
\begin{gathered}
\frac{\partial \braket{F}_d}{\partial \varphi_{+}}=-\frac{1}{2 \widetilde{q}} \sqrt{\frac{h \widetilde{q}}{2}} \frac{1}{\mathcal{V}}\sum_{k\alpha}\left\langle \braket{s_{k \alpha} y_{k \alpha}}\right\rangle_d+2\varphi_{-}=0 \\
\frac{\partial \braket{F}_d}{\partial \varphi_{-}}=-\frac{h}{4\mathcal{V}} \sum_{k\alpha}\braket{\braket{ s_{k \alpha}^2}}_d+\frac{1}{4 \beta \widetilde{q}} \sqrt{\frac{h \widetilde{q}}{2}}\frac{1}{\mathcal{V}} \sum_{k\alpha}\braket{\left\langle s_{k \alpha} y_{k \alpha}\right\rangle}_d+2\varphi_{+}=0 \\
\frac{\partial \braket{F}_d}{\partial \Delta}=-\frac{1}{\mathcal{V}}\sum_k\braket{\left\langle s_{k x}\right\rangle}_d+\frac{2}{\bar{U}} \Delta=0
\end{gathered}
\end{equation}
where we define 
\begin{equation}
\begin{aligned}
& \braket{\braket{ A(s_{k}^{(z)},s_{k}^{( \alpha)})}}_d \\
& =\int \frac{d \bs{y_{k}}}{2 \pi} \exp{\left[-\frac{\left(\bs{y_{k}}\right)^2}{2}\right]}\left(\frac{1}{Z_{0 k}}\right) \int\left[A(s_{k}^{(z)},s_{k}^{( \alpha)})\right] D [S_k] e^{-\beta \left[ \left( -\xi_k s_{k}^{(z)}-\sum_{\alpha} \frac{1}{4}h\varphi_{-}(s^{(\alpha)}_{k})^2 -\sum_{\alpha} \left(\sqrt{\frac{h\widetilde{q}}{2}}y_{k\alpha}+\Delta\delta_{\alpha x}\right)s^{(\alpha)}_{k} \right) \right]}
\end{aligned}
\end{equation}

In the limit $\beta \to \infty$ , we obtain 
\begin{equation}
\begin{gathered}
-\frac{1}{4 \varphi_+} \sqrt{\frac{h \varphi_+}{2}}\frac{1}{\mathcal{V}} \sum_k\braket{ \braket{\bs{s^{(\perp)}_{k}} \cdot \bs{y_{k }}}}_d+\varphi_{-}=0 \\
-\frac{h}{8\mathcal{V}} \sum_k\braket{ \braket{(\bs{s^{(\perp)}_{k}})^2}}_d+\varphi_{+}=0 \\
-\frac{1}{\mathcal{V}}\sum_k\braket{ \braket{\bs{s^{(\perp)}_{k}} \cdot \bs{\hat{x}} }}_d+\frac{2}{\bar{U}} \Delta=0.
\end{gathered}
\end{equation}

We next consider the averages $\braket{\braket{\cdots}}_d$. We note that $\braket{\cdots}$ is an expectation value weight over the effective free energy $f_{eff}(\vec{s_k},\bs{y_{k}})=-\xi_k s_{k}^{(z)}-\sum_{\alpha} \frac{1}{4}h\varphi_{-}(s^{(\alpha)}_{k})^2 -\sum_{\alpha} (\sqrt{\frac{h\varphi_{+}}{2}}y_{k\alpha}+\Delta\delta_{\alpha x})s^{(\alpha)}_{k}$. In the $\beta \to \infty$ limit, we can evaluate the integral over $D[S_k]$ by replacing it with the saddle point value $\braket{ A(s_{km}^{(z)},s_{km}^{( \alpha)})}$, where $s^{(z)}_{km}$ and $s^{(\alpha)}_{km}$ are obtained by minimizing 
\begin{equation}
\begin{aligned}
    f_{eff}(\vec{s_k},\bs{y_{k}}) &= -\xi_k s^{(z)}_{k} - \bs{s^{(\perp)}_k} \cdot (\sqrt{\frac{h\varphi_{+}}{2}}\bs{y_{k}}+\Delta \bs{\hat{x}})-\frac{1}{4}h\varphi_-(\bs{s_k^{\perp}})^2 \\
    &= -\xi_k x_k -\left|\sqrt{\frac{h\varphi_{+}}{2}}\bs{y_{k}}+\Delta \bs{\hat{x}}\right|\sqrt{1-x_k^2} \cos{\phi}-\frac{1}{4}h\varphi_-(1-x_k^2)
\end{aligned}
\label{B12}
\end{equation}
where $\bs{s^{(\perp)}_k}=(s^{(x)}_{k},s^{(y)}_{k})$ and $\phi$ is the angle between the pairing field $\bs{B}=\sqrt{\frac{h\varphi_{+}}{2}}\bs{y_{k}}+\Delta \bs{\hat{x}}$ and $\bs{s^{(\perp)}_k}$. We have reparametrized $f_{eff}(\vec{s_k},\bs{y_{k}})$ by $s^{(z)}_{k}=\cos{\theta}=x_k$ and $(\bs{s^{(\perp)}_k})^2=\sin^2{\theta}=1-x_k^2$ in the second line of Eq.\ (\ref{B12}). $f_{eff}(\vec{s_k},\bs{y_{k}})$ is minimized when $\cos{\phi}=1$ and $x_k=x_{km}$ where $x_{km}$ satisfies
\begin{equation}
    -\xi_k+\frac{x_{km}B}{\sqrt{1-x_{km}^2}}+\frac{1}{2}h\varphi_-x_{km}=0
    \label{xi-x}
\end{equation}
where $B=|\bs{B}|=\left|\sqrt{\frac{h\varphi_{+}}{2}}\bs{y_{k}}+\Delta \bs{\hat{x}}\right|$. \\ \\

To compute the sum over $k$, we convert it to an integral over single particle energy $\xi_k$, i.e. $\frac{1}{L^d}\sum_k(\cdots) \to N(0)(\int_{-\omega_D}^{\epsilon_c}(\cdots)d\xi+\int_{\epsilon_c}^{\omega_D}(\cdots)d\xi)$, where $N(0)$ is the density of states at the Fermi surface and $\epsilon_c\sim 1/(N(0)L^d)$ is the energy cutoff associated with a finite sized grain. The approximation is valid as long as the grain size is not too small. Using Eq.(\ref{xi-x}), we obtain 
\begin{equation}
d \xi=\left\{\frac{B}{\left(1-x_m^2\right)^{\frac{3}{2}}}+\frac{1}{2}h \varphi_{-}\right\} d x_m.
\label{int_var}
\end{equation}
 the integral over $\xi$ or $x_m$ can be evaluated analytically in the $\omega_D \to \infty$ limit and we obtain the self-consistent equations at $T=0$
\begin{equation}
    \varphi_- = \frac{1}{4 \varphi_{+}} \sqrt{\frac{h \varphi_{+}}{2}}N(0) \int \frac{d\bs{y}}{2\pi} e^{-\bs{y}^2/2} \frac{\bs{B}\cdot \bs{y}}{B} \left[B\ln{\frac{1-x_c}{1+x_c}}+2B\ln{\frac{2\omega_D}{B}}+\frac{h\varphi_-}{2}(-x_c\sqrt{1-x_c^2}-\arcsin{x_c}+\frac{\pi}{2})\right]
    \label{phim}
\end{equation}
\begin{equation}
    \varphi_+ = \frac{h}{8} N(0) \int \frac{d\bs{y}}{2\pi} e^{-\bs{y}^2/2} \left[-2B\arcsin{x_c}+\pi B + h\varphi_-(-x_c+x_c^3/3) + \frac{2h\varphi_-}{3}\right]
    \label{phip}
\end{equation}
\begin{equation}
    \Delta = \frac{\bar{U} N(0)}{2} \int \frac{d\bs{y}}{2\pi} e^{-\bs{y}^2/2} \frac{\bs{B} \cdot \bs{\hat{x}}}{B} \left[B\ln{\frac{1-x_c}{1+x_c}}+2B\ln{\frac{2\omega_D}{B}}+\frac{h\varphi_-}{2}(-x_c\sqrt{1-x_c^2}-\arcsin{x_c}+\frac{\pi}{2})\right]
    \label{Delta}
\end{equation}
where $x_c \sim \frac{\epsilon_c}{B+\frac{1}{2}h\varphi_-}$ when $x_c$ is small.
The integral over $d\bs{y}$ cannot be evaluated analytically and the above self-consistent equations have to be solved numerically to obtain the order parameters $\varphi_+$, $\Delta$ and $\varphi_-$. 

\section{Stability of the replica-symmetric theory}
\label{Stability}
In this Appendix we show that the replica-symmetric theory is stable with respect to replica symmetry breaking terms. To see that we write $q_{ij}=q+\delta q_{ij}$ where $q$ is the replica-symmetric solution and $\delta q_{ij}$ is a small correction that breaks replica symmetry. We now examine the $n$-replica partition function
\begin{equation}
    \braket{Z}_d = \text{Tr}e^{-\beta H_R^{(n)}}
\end{equation}
where Tr denotes the sum over all possible spinor configurations and the disorder-average over $\bs{y}_k$. By replacing $q_{ij}$ by ${q+\delta q_{ij}}$, we have in the large $\beta$ limit 
\begin{equation}
\begin{gathered}
H_R^{(n)} = \sum_{i k} -\xi_k s_{k i}^{(z)}-\sum_{i k} \bs{s}_{k i}^{\perp} \cdot \left(\Delta \bs{\hat{x}}+\sqrt{\frac{h}{2} \varphi_{+}} \bs{y}_k\right)-\frac{1}{4} \sum_{i k}h \varphi_{-} \bs{s}_{k i}^{\perp} \cdot \bs{s}_{k i}^{\perp}-\frac{h}{4} \sum_{i \neq j, k} \delta q_{i j} \bs{s}_{k i}^{\perp} \cdot \bs{s}_{k j}^{\perp} \\
+\sum_k \frac{\left(\bs{y}_k\right)^2}{2 \beta}+\frac{n^2 q^2}{\beta}+n \frac{\Delta^2}{U_0}+2 n \varphi_{+} \varphi_{-}+\frac{1}{\beta} \sum_{i \neq j} {\delta q_{i j}}^2
\end{gathered}
\end{equation}
We note that as in the replica-symmetric solution the partition function $\braket{Z^n}_d$ is dominated by configurations $\left\{\bs{s}_{k i}\right\}$ that minimizes $H_R^{(n)}$ at zero temperature, i.e. we have to minimize
\begin{equation}
H_{e f f}^{(n)}(s)=-\sum_{i k} \xi_k s_{k i}^{(z)}-\sum_{i k} \bs{s}_{k i}^{\perp} \cdot\left(\Delta \bs{\hat{x}}+\sqrt{\frac{h}{2} \varphi_{+}} \bs{y}_k\right)-\frac{1}{4} \sum_{i k}h \varphi_{-} \bs{s}_{k i}^{\perp} \cdot \bs{s}_{k i}^{\perp}-\frac{h}{4} \sum_{i \neq j, k} \delta q_{i j} \bs{s}_{k i}^{\perp} \cdot \bs{s}_{k j}^{\perp}
\end{equation}
with respect to $\bs{s}_{k i}$. Writing
\begin{equation}
H_{e f f}^{(n)}(s)=H_0^{(n)}(s)+V^{(n)}(s)
\end{equation}
where
\begin{equation}
H_0^{(n)}(s) =  \sum_{ik}\left(-\xi_k s_{k i}^{(z)}-h \gamma_0 \bs{s}_{k i}^{\perp} \cdot \bs{s}_{k i}^{\perp}\right)-\sum_{ik} \bs{s}_{k i}^{\perp} \cdot \bs{B}
\end{equation}
and
\begin{equation}
V^{(n)}(s)=-\frac{h}{4} \sum_{i \neq j,k} q_{i j} \bs{s}_{k i}^{\perp} \cdot \bs{s}_{k j}^{\perp}
\end{equation}
Where $ \gamma_0=\frac{1}{4} \varphi_{-}, \bs{B}=\Delta \bs{\hat{x}}+\sqrt{\frac{h}{2} \varphi_{+}} \bs{y}_k$ 
and using the notation in Appendix \ref{MFE}, we have
\begin{equation}
H_0^{(n)}(s) = \sum_{ik} \left(-\xi_k x_{k i}-h \gamma_0\left(1-x_{k i}^2\right)-\sqrt{\left(1-x_{k i}^2\right)} B\right).
\end{equation}
Next we expand the solution in powers of $\delta q_{ij}$, i.e. we write $x_{k i}=x_k+w_{k i}$ with $x_k$ corresponds to the replica-symmetric solution and $w_{ki}$ represents the replica symmetry breaking contribution. \\ \\
We obtain
\begin{equation}
\begin{aligned}
 H_0^{(n)}(s) &= \sum_{ik} \left(-\xi_k\left(x_k+w_{k i}\right)-h \gamma_0\left(1-\left(x_k+w_{k i}\right)^2\right)-\sqrt{\left(1-\left(x_k+w_{k i}\right)^2\right)} B\right. \\
& \approx \left[\sum_{ik} \xi_k x_k-h \gamma_0\left(1-x_k^2\right)-\sqrt{1-x_k^2} B+h \gamma_0\left(w_{k i}\right)^2+\frac{\left(w_{k i}\right)^2 B}{2\left(1-x_k^2\right)^{\frac{3}{2}}}\right]
&
\end{aligned}
\end{equation}
and
\begin{equation}
\begin{aligned}
V^{(n)}(s)&=-\frac{h}{4} \sum_{i \neq j,k} q_{i j} \sqrt{1-\left(x_k+w_{k i}\right)^2} \sqrt{1-\left(x_k+w_{k j}\right)^2}\\
& \approx -\frac{h}{4} \sum_{i \neq j,k} q_{i j}\left(1-x_k^2\right) \times\left(1-\frac{w_{k i} x_k}{\left(1-x_k^2\right)}-\frac{w_{k j} x_k}{\left(1-x_k^2\right)}-\frac{\left(w_{k i}\right)^2}{2\left(1-x_k^2\right)^2}-\frac{\left(w_{k j}\right)^2}{2\left(1-x_k^2\right)^2}+\frac{w_{k i} w_{k j} x_k^2}{\left(1-x_k^2\right)^2}
\right) \\
\end{aligned}
\end{equation}
\\ 
valid to order $(w_{k\alpha})^2$. Minimizing the total energy with respect to $w_{k i}\sim w_{k i}^{(1)}+w_{k i}^{(2)}$ and expanding the result in powers of $\delta q_{ij}$, we obtain
\begin{equation}
H_{eff}^{(n)}(s) = H_{eff}^{(n 0)}(s)+H_{eff}^{(n 1)}(s)+H_{eff}^{(n 2)}(s)+\cdots
\end{equation}
with \begin{equation}
H_{eff}^{(n 0)}(s)=n \sum_{i,k}\left(-\xi_k x_k-h \gamma_0\left(1-x_{k}^2\right)-\sqrt{1-x_k^2} B\right) 
\end{equation}
\begin{equation}
H_{eff}^{(n 1)}(s)=-\frac{h}{4} \sum_{i \neq j,k} q_{i j}\left(1-x_k^2\right)=0
\end{equation}
and
\begin{equation}
\begin{aligned}
H_{eff}^{(n 2)}(s)&=-\sum_{i,k} \frac{\left(h x_k\right)^2}{8\left(2 h \gamma_0+\frac{B}{\left(1-x_k^2\right)^{\frac{3}{2}}}\right)} \sum_{j l} \delta q_{i j} \delta q_{i l} \\
&=-z\sum_{i,j,l} \delta q_{i j} \delta q_{i l}
\end{aligned}
\end{equation} with $z\sim\frac{N(0)h^2}{12}(1-x_c^3)>0$,
where $x_c \sim \frac{\epsilon_c}{B+\frac{1}{2}h\varphi_-}$ and we have performed the sum over $k$ using the trick employed in Appendix \ref{MFE}(see Eq.\ (\ref{int_var})). \\ \\ \\
The replica-symmetric solution is stable if the quadratic form $H_{R}^{(n 2)}(s)$ is at least positive semi-definite. To check this, we need to show that all eigenvalues of the Hessian matrix are all non-negative in $n\to 0$ limit when $\beta\rightarrow\infty$. We find that for a general $n$, the eigenvalues of the Hessian are given by $(-2n+2)z$, $-(n-2)z$ and $0$. 
In the $n \to 0$ limit, all eigenvalues are non-negative. Our result is consistent with the more general result given in \cite{de1978stability}.

\section{Some details in calculation of physical obervables}
\label{PO}
In this Appendix, we supply some calculation details for the disorder-average single-particle density of states and the superfluid density.

\subsection{Single-particle density of states}
The single particle density of states can be obtained by summing over $k$ of the imaginary part of the retarded Green function, i.e. 
\begin{equation}
    D(\omega)=-\frac{1}{V\pi} \sum_k \text{Im} G^R(k,\omega).
\end{equation}
Using Eqs.(\ref{single dos}) and (\ref{uv}) in the main text, we obtain
\begin{equation}
\begin{aligned}
& \quad
\braket{\braket{D(\omega)}}_d \\
    &=\frac{1}{2V}  \sum_k \int \frac{\text{d}\bs{y_k}}{2\pi} \exp{\left[-\bs{y_k}^2/2\right]}[(1+x_{km})\delta(\omega-\xi_k/x_{km}-U/2)+(1-x_{km})\delta(\omega+\xi_k/x_{km}+U/2)]  \\ 
    &=\frac{1}{2V} N(0) \int \frac{\text{d}\bs{y_k}}{2\pi} \exp{\left[-\bs{y_k}^2/2\right]}\int_{-\sqrt{1-\frac{B^2}{\omega_D^2}}}^{\sqrt{1-\frac{B^2}{\omega_D^2}}} \text{d}x_m (\frac{B}{(1-x_m^2)^{\frac{3}{2}}}+\gamma) [(1+x_{m})\delta(\omega-\xi /x_{m}-U/2)+(1-x_{m})\delta(\omega+\xi/x_{m}+U/2)],
    \label{dos}
\end{aligned}
\end{equation}
where we have use Eq.(\ref{int_var}) and approximate $\sum_k(\cdots) \to N(0)\int_{-\omega_D}^{\omega_D}(\cdots)d\xi$. \\ \\
Eliminating the $\delta$ function we obtain
\begin{equation}
\braket{\braket{D(\omega)}}_d=\frac{N(0)}{V}\int \frac{\text{d}\bs{y_k}}{2\pi} \exp{\left[-\bs{y_k}^2/2\right]} \left[\frac{1}{\sqrt{(|\omega|-\gamma)^2-B^2}}\left((|\omega|-\gamma)+\frac{B^2\gamma}{(|\omega|-\gamma)^2}\right)\right]\theta(|\omega|-\gamma-B),
\end{equation}
where $\gamma = \frac{1}{2}( h\varphi_-+U)$ and $B=|\bs{B}|=\left|\sqrt{\frac{h\varphi_{+}}{2}}\bs{y_{k}}+\Delta \bs{\hat{x}}\right|$. The integral is evaluated numerically.

\subsection{Superfluid density}
We next consider the derivation of Eq.(\ref{rho}). To derive the function $L(\xi,\xi')$, we start with the imaginary-time current-current response function in the frequency space
\begin{equation}
\begin{aligned}
    &\quad \quad \chi_j^{\alpha \beta}(\bs{r},\bs{r'},i\omega_n) \\
    &= -\int_0^\beta \text{d}\tau e^{i\omega_n \tau} \braket{T_{\tau}j^\alpha(\bs{r},\tau)j^\beta(\bs{r'},0)} \\
    &\sim -\int_0^\beta \text{d}\tau e^{i\omega_n \tau} \sum_{nm\sigma} \sum_{pqs} \{\braket{T_\tau c^\dag_{m\sigma}(\tau) c^\dag_{ps}(0)} \braket{T_\tau c_{qs}(0)c_{n\sigma}(\tau)}-\braket{T_\tau c_{n\sigma}(\tau) c^\dag_{ps}(0)} \braket{T_\tau c_{qs}(0) c^\dag_{m\sigma}(\tau)}  \}  J_{mn}^\alpha(\bs{r}) J_{pq}^\beta(\bs{r'}) \\
    &\sim \int d\xi \int d\xi'  L(\xi,\xi',i\omega_n) \sum_{nm} \delta(\xi-\xi_n) \delta(\xi'-\xi_m) J_{mn}^\alpha(\bs{r}) J_{nm}^\beta(\bs{r'}) \\
    &\xrightarrow{i\omega_n \to 0} \int d\xi \int d\xi'  L(\xi,\xi') \operatorname{Re\sigma }(\xi-\xi',\bs{r},\bs{r'}),
\end{aligned}
\end{equation}
where 
\begin{equation}
    L(\xi,\xi',i\omega_n) = \frac{1}{2} \frac{(E+E')(E E'-\Delta_{\xi} \Delta_{\xi'}-\xi \xi')+i\omega_n(E' \xi-E \xi')}{E E'(E+E'-i\omega_n)(E+E'+i\omega_n)}
\end{equation}
and
\begin{equation}
L\left(\xi, \xi^{\prime}\right)=\frac{1}{2}\frac{E E'-\Delta_{\xi}^* \Delta_{\xi^{\prime}}-\xi \xi^{\prime}}{E E'\left(E+E'\right)}
\end{equation}
with $E=\sqrt{\xi^2+|\Delta_ \xi|^2}$. 
The real part of the conductivity tensor at normal state is given by
\begin{equation}
\operatorname{Re\sigma }(\xi-\xi',\bs{r},\bs{r'})\sim \sum_{nm} \delta(\xi-\xi_n) \delta(\xi'-\xi_m) J_{mn}^\alpha(\bs{r}) J_{nm}^\beta(\bs{r'}),
\end{equation}
where $\bs{J}_{nm}(\bs{r})=\frac{ie}{2m_e}\left(\phi^*_m(\bs{r})\nabla \phi_n(\bs{r})-\phi_n(\bs{r})\nabla \phi_m^*(\bs{r})\right)$. 
We employ a trick by noting that $\rho_s=0$ in the normal state and we can write\cite{ma1985localized}
\begin{equation}
\begin{aligned}
\rho_s &= 
 \frac{2}{ \pi c} \int d\xi \int d\xi'\left[L\left(\xi, \xi^{\prime}\right)-L\left(\xi, \xi^{\prime}\right)_{\Delta=0}\right] \operatorname{Re\sigma }\left(\xi-\xi^{\prime}\right)
\end{aligned}
\label{SD}
\end{equation}
\\ \\
where 
\[  \operatorname{Re\sigma} (\xi-\xi^{\prime})=\int d^dr\braket{\operatorname{Re\sigma }(\xi-\xi',\bs{r},\bs{r'})}_d. \]


For a localized system in 2 dimensions \cite{vollhardt1980diagrammatic},
\begin{equation}
\operatorname{Re\sigma} (\xi-\xi^{\prime})\sim \frac{n e^2 \tau}{m}\left(\frac{(\xi-\xi')^2}{\tau^2 \omega_0^4+(\xi-\xi')^2}\right)
\label{cond}
\end{equation}
where $\omega_0$ is the characteristic frequency(energy) below which all single particle states are localized. Notice that $\omega_0$ is given by $\omega_0=\sqrt{2E_F/md}L^{-1}$ in $d$ dimensions \cite{vollhardt1980diagrammatic} and thus $\omega_0 = \frac{v_F}{\sqrt{2}L}$ for $d=2$. Using $h^2\sim \frac{\sigma_U^2D_I^2}{L^{4}}+\frac{U_0^2}{L^{2}}$ and $L \sim l e^{\frac{\pi}{2}k_Fl}$ in 2 dimensions, where $l$ is the mean free path, we can estimate $l$ by solving the above relation and determine $\tau=l/v_F$.

\end{document}